\documentclass[12pt,a4paper,thmsb]{article}
\usepackage{eurosym}
\usepackage{amsmath}
\usepackage{amssymb}
\usepackage{amsfonts}
\usepackage{color}
\usepackage{comment}

\newtheorem{theorem}{Theorem}

\newtheorem{proposition}{Proposition}

\newtheorem{example}{Example}
\newenvironment{proof}[1][Proof]{\noindent\textbf{#1.} }{\ \rule{0.5em}{0.5em}}

\long\def\comment#1{}

\begin{document}
\renewcommand{\thefootnote}{\fnsymbol{footnote}}

\title{Integrating equity and productivity in health evaluation\footnotemark[1]}

\author{Kristian S. Hansen\footnotemark[2] 
\and Juan D. Moreno-Ternero\footnotemark[3]
\and Lars P. \O sterdal\footnotemark[4]}
\footnotetext[1]{ 
We thank Odd Rune Straume, Ramses Abul Naga, as well as audiences at Universidad de Montevideo, Universidad de Chile, Kyoto University, and University of Granada, for helpful comments and suggestions. 
Financial support from the National Research Centre for the Working Environment (NFA), Copenhagen, Denmark, the Spanish Agencia Estatal de Investigaci\'on (AEI) via grant
PID2023-146364NB-I00 funded by MCIU/AEI/10.13039/501100011033 and FSE+, and the Independent Research Fund Denmark (grant no. 4260-00050B), is gratefully acknowledged.}
\footnotetext[2]{National Research Centre for the Working Environment (NFA), Copenhagen, Denmark. e-mail: krh@nfa.dk}
\footnotetext[3]{Department of Economics, Universidad Pablo de Olavide. e-mail: jdmoreno@upo.es}
\footnotetext[4]{Department of Economics, Copenhagen Business School. e-mail: lpo.eco@cbs.dk  }

\date{
\today}

\maketitle

\renewcommand{\thefootnote}{\arabic{footnote}}

\begin{abstract}

This paper develops a unified framework for evaluating health outcomes that jointly incorporates equity and productivity. Extending beyond traditional QALYs, PALYs, and the more recent PQALYs, we introduce a broader class of evaluation functions that integrate equity- and productivity-sensitive conditions. By imposing several normative criteria, including independence from measurement scales and Pigou-Dalton transfer principles, we obtain tractable power-form representations. In balancing equity and efficiency, the framework provides a coherent foundation for assessing interventions in contexts where both health and productive capacity are at stake.
\end{abstract}

\medskip{}

\medskip{}  \medskip{}

\newpage

\section{Introduction}
Health interventions affect not only the level and distribution of health, but also the productive capacity of individuals and populations. In many policy domains, ranging from health care to occupational health and safety, resource allocation decisions therefore involve trade-offs between improving average health, protecting vulnerable groups, and sustaining productivity. Conventional evaluation tools used in health economics typically focus on health outcomes alone and provide only limited guidance on how to balance these objectives when they conflict. This creates a need for outcome measures that can systematically incorporate both equity concerns and productivity effects within a single, coherent framework.

Standard approaches to health evaluation rely on summary measures of population health that aggregate longevity and quality of life into a single index. Models such as Quality-Adjusted Life Years (QALYs) and Disability-Adjusted Life Years (DALYs) have been widely adopted to capture these outcomes (e.g., Murray, 1996; Drummond et al., 2015; Neumann et al., 2016; GBD 2021 Diseases and Injuries Collaborators, 2024). Although valuable, these metrics have been criticized for overlooking broader societal concerns, particularly those related to equity (e.g., Broome, 1993) and the distribution of benefits across population groups. In response, a substantial body of work has developed equity-sensitive health evaluations, adjusting conventional measures to prioritize vulnerable individuals and align with principles of distributive justice (e.g., Wagstaff, 1991; Williams, 1997; Adler, 2012, 2025; Cookson et al., 2020; Da Costa et al., 2024; Robson et al., 2024, 2025b).

From a different vantage point, scholars have increasingly emphasized the role of productivity as an integral component of health outcomes, rather than treating it only as a secondary or indirect consideration. A wide literature shows how poor health impairs work performance and generates substantial social costs (e.g., Zhang et al., 2011; Lohaus and Habermann, 2019; Elsler et al., 2017). In response, outcome-oriented approaches such as Productivity-Adjusted Life Years (PALYs) have been proposed to incorporate productivity directly into health metrics, reflecting a growing consensus to evaluate interventions on both productivity and health outcomes (e.g., Mattke et al., 2007; Ademi et al., 2021, 2025; Premji and Griffin, 2024). In Hansen et al. (2024), we extended this line of work to propose Productivity- and Quality-Adjusted Life Years (PQALYs), which integrate productivity and quality of life into a single metric. This framework recognizes that health not only influences longevity and well-being but also shapes individuals' capacity to contribute economically. In doing so, PQALYs broaden the scope of evaluations beyond QALYs and DALYs, allowing analysts to capture both health-related quality adjustments and productivity considerations within the same outcome measure. The approach is particularly relevant for domains such as occupational health and safety (OHS), where the dual impacts of interventions on well-being and workforce participation are especially pronounced (e.g., Tompa et al., 2008; Steel et al., 2018).

This paper develops a unified framework for health evaluation that integrates equity and productivity dimensions within an axiomatic setting. Our approach generalizes the PQALY proposal in Hansen et al. (2024) by embedding it within a broader class of evaluation functions that reflect fairness and economic sustainability. The framework connects classical tools from welfare economics (e.g., Moulin, 1988) with recent axiomatic advances in population health analysis (e.g., Hougaard et al., 2013; Moreno-Ternero et al., 2023), thereby providing a new normative foundation for assessing interventions in contexts where the equity-sensitive interplay between health and productive capacity is crucial. 

To achieve this, we outline a set of core axioms for evaluating population health distributions that are natural in this setting, incorporating Pigou-Dalton principles to capture equity considerations while imposing only minimal structural assumptions. From these foundations, we derive several formulations of evaluation functions based on Healthy Productive Years Equivalents (HPYEs), which extend beyond traditional QALYs and HYEs by capturing the joint effects of health status and productive capacity.\footnote{Healthy-Years Equivalents (HYEs) were introduced by Mehrez and Gafni (1989) as an alternative to QALYs. Moreno-Ternero et al. (2023) extend the notion to account for reference lifetimes, whereas Hansen et al. (2024) do so to account for productivity.} A central feature of our analysis is the use of measurement-scale independence assumptions, which yield tractable power-form measures sensitive to disparities in health and productivity. Time scale independence reflects the view that lifetime length enters the evaluation only through relative duration. Accordingly, rescaling all lifetimes by a common factor, for example changing the unit of time from years to months, does not affect the social ranking of distributions.

%A central feature of our analysis is the use of measurement-scale independence assumptions, which yield tractable power-form measures sensitive to disparities in health and productivity. Time scale independence reflects the assumption that lifetime length enters social evaluation only through \textit{relative duration}. Under this view, rescaling all lifetimes by a common factor, for example, changing the unit of time from years to months, does not alter the socially relevant comparisons among individuals. Adopting this assumption commits the analysis to a particular view of time: what matters is relative lifetime length, not the numerical values induced by a specific temporal scale. The characterization results that rely on time scale independence should therefore be understood as conditional on this interpretation; alternative views that attach significance to particular temporal units, life stages, or calendar thresholds would motivate different axioms, which would lead to different evaluative criteria.

As a preview of our results, we first note that our framework is grounded in a characterization of equity-sensitive preferences represented by generalized HPYE evaluation functions with concave aggregators (Theorem 1). Imposing an additional time scale independence condition at full health and maximal productivity yields the equity-sensitive power HPYE evaluation function (Theorem 2), whereas a more general time scale independence condition leads to an equity-sensitive productivity- and quality-adjusted power lifetime evaluation function (Theorem 3). As limiting cases, imposing productivity or health independence, respectively, recovers the equity-sensitive power QALY or equity-sensitive power PALY evaluation functions. Moving beyond these broad formulations, we derive an equity-sensitive power PQALY evaluation function that fully integrates quality-of-life and productivity adjustments into the life-year metric (Theorem 4). We also obtain two hybrid specifications (power QALY-PALY and power QALY-PQALY) that introduce partial productivity or quality-of-life sensitivity, respectively, thereby bridging the pure power QALY metric and its productivity (power PALY) or productivity-adjusted (power PQALY) counterparts (Theorems 5 and 6). Finally, imposing a productivity scale independence axiom alongside a Pigou-Dalton principle for productivity yields an equity-sensitive bi-power PQALY function with distinct inequality-aversion parameters for lifetime and productivity (Theorem 7).

%A final clarification concerns the relationship between health and productivity. 
Empirically, a substantial literature shows that improvements in health often raise productivity, through channels such as cognitive performance, physical capacity, and labor supply (see, among many others, Strauss and Thomas, 1998; Currie and Madrian, 1999; Bloom et al., 2004; Weil, 2007). The relationship may also run in the opposite direction, as adverse working conditions, high work intensity, and long hours can deteriorate mental and physical health, with implications for subsequent labor market outcomes and diminishing productivity returns (e.g., Pencavel, 2018; Belloni et al., 2022; Jolivet and Postel-Vinay, 2025).
In the framework we consider in this paper, such effects are naturally represented as joint changes in health and productivity within an individual profile, rather than as changes in a single attribute. 
Treating health and productivity as distinct variables places no restriction on their empirical interaction, but instead keeps the evaluation problem explicit about how health and productivity gains are normatively weighted. This separation is essential for our axiomatic results, which characterize equity-sensitive aggregation rules and allow for different degrees of inequality aversion with respect to health and productivity.

Finally, we emphasize that the present paper provides the first full axiomatic characterizations of equity-sensitive aggregation in a framework where individual equivalents are defined jointly over health and productivity. Hansen et al. (2024) showed how profiles of health states, productivity, and longevity can be represented by HPYEs and aggregated in meaningful ways, including a characterization of unweighted PQALY aggregation as a unified productivity- and quality-adjusted life-year metric. However, the evaluation functions characterized there ultimately rely on utilitarian aggregation of individual equivalents and thus abstract from distributive trade-offs. Here, we move beyond this framework by introducing explicit distributive principles such as Pigou-Dalton transfer principles for lifetime and productivity, together with measurement-scale independence requirements, that sharply constrain admissible departures from utilitarianism. These axioms rule out unweighted aggregation and lead to richer representation results. 
Rather than fixing a single equity correction, the theory yields structured families of admissible evaluation criteria that jointly evaluate health and productivity, accommodate inequality aversion within and across dimensions, and allow distinct normative weights on the two dimensions.
%Rather than fixing a single equity correction, the theory delivers structured families of admissible evaluation criteria that jointly evaluate productivity and health, accommodate inequality aversion within and across dimensions, and allow distinct normative weights on health and productivity.
%Rather than fixing a single equity correction, the theory delivers structured families of admissible evaluation criteria that integrate productivity and quality of life while accommodating inequality aversion within and across dimensions, including distinct normative weights on health and productivity.

The rest of the paper is organized as follows. In Section 2, we introduce the preliminaries of our analysis. 
In Section 3, we characterize broad classes of equity-sensitive evaluation functions. Section 4 characterizes focal members of those classes. Section 5 discusses limitations of our analysis, as well as future research directions. Proofs are relegated to the Appendix.

\section{Preliminaries}\label{sec:prelim}
\subsection{Definitions}
We build on the model introduced by Hansen et al. (2024) for our analysis. 
%and direct readers to that paper for a more in-depth discussion. 
Let a population consisting of $n$ individuals be identified with the set $N=\{1,...,n\}$. Each individual $i \in N$ is described by a profile, formalized by a triple $d_i=(a_i, p_i, t_i)$, where $a_i\in A$ is a health state, $p_i \in  [0, 1]$ is the productivity level, and $t_i \in \Bbb{R}_+ = [0, \infty)$ is time. 

Note that the set of health states $A$ is broadly defined without mathematical restrictions and is conceived to capture the complex and multifaceted nature of health. As observed by Pliskin et al. (1980), these health states can be interpreted as representative conditions over the entire lifespan. We do assume that there exists a health state $a_\ast\in A$, referred to as `full health', which is considered at least as good as any other health state. Productivity is defined to reflect an individual’s overall productive capacity, encompassing aspects such as work attendance, on-the-job performance, and contributions to economic, household, or community activities. Note that productivity is expressed in relative terms, such that $p_i= 1$ represents maximal productivity, while $p_i= 0$ represents null productivity. %Note also that our model does not exclude from the outset cases in which health and productivity are not correlated. But the remaining values for productivity are perfectly ranked according to the standard Euclidean dimension. 
In our framework, time $t_i$ denotes individual $i$'s remaining lifetime within the analysis period, which may refer to years until retirement, the end of life, or another relevant horizon, allowing the model to be used in both working-age and full-lifetime evaluations.

Let $d=(d_1,...,d_n)$ denote a distribution of individual profiles, as described above, and let $D$ denote the set of possible distributions. 
The preferences of a policymaker (or social preferences) over distributions are given by $\succsim$.  
A distribution evaluation function (\textit{evaluation function}, in short) is a real-valued function $E: D \rightarrow \mathbb{R}$. We say that $E$ represents $\succsim$ if higher values of $E$ correspond to more preferred distributions.\footnote{ Formally, $E[d]%=E[(d_1,...,d_n)] =E[(d'_1,...,d'_n)]
%if it preserves the ordering of distributions in the sense that higher function values reflect preferred distributions.
\ge E[d'] \Leftrightarrow d \succsim d'.$ }
When applied in economic evaluation of health care or OHS interventions, $E$ may serve as an effect measure.

Some instances of distributions of individual profiles are presented in the following example. They will be used later to illustrate the implications of different axioms and evaluation functions.

\begin{example} %\textbf{This is the same example as in Hansen et al. (2024). Should we update it?} 
    
    Consider the following two distributions, %($\Omega=(d^{\Omega}_1,d^{\Omega}_2,d^{\Omega}_3,d^{\Omega}_4,d^{\Omega}_5)$, $\Phi=(d^{\Phi}_1,d^{\Phi}_2,d^{\Phi}_3,d^{\Phi}_4,d^{\Phi}_5)$), 
    involving five individuals each (which could represent five equally sized groups). 
    
    %In summary, 
    $$d^{\Omega}=\left(\begin{array}{ccc}d^{\Omega}_1\\ 
    d^{\Omega}_2\\
    d^{\Omega}_3\\ 
    d^{\Omega}_4\\
    d^{\Omega}_5 \end{array}\right) 
    =
    \left(\begin{array}{ccc}a_\ast & 1 & 40\\ 
    a_\ast & 0.5 & 5\\
    a_\ast & 0 & 20\\ 
    a_\ast & 0.5 & 5\\
    a_\ast & 0 & 0 \end{array}\right); \quad 
    d^{\Phi}=\left(\begin{array}{ccc}d^{\Phi}_1\\ 
    d^{\Phi}_2\\
    d^{\Phi}_3\\ 
    d^{\Phi}_4\\
    d^{\Phi}_5 \end{array}\right) 
    =
    \left(\begin{array}{ccc}a_\ast & 1 & 40\\ 
    a & 0 & 10\\
    a & p & 10\\ 
    a & 0 & 10\\
    a & 0 & 0\end{array}\right).$$
    %\[\Omega=\left( \begin{array}{ccc}a_\ast & 1 & 40\\ a_\ast & 0.5 & 40\\a_\ast & 0 & 40\\ a_\ast & 0.5 & 10\\a_\ast & 0 & 40 \end{array}%\right) . \]
    %\[\Phi=\left( \begin{array}{ccc}a_\ast & 1 & 40\\ a & 1 & 40\\a & 0.5 & 40\\ a & 0.5 & 10\\a & 0 & 40 \end{array}%\right) . \]
\medskip

    In the first distribution ($d^{\Omega}$), all individuals are experiencing full health. The first individual has maximal productivity as well as forty years of lifetime, i.e., $d^{\Omega}_1=(a_\ast,1,40)$. The second individual has $50\%$ productivity and five years of lifetime, i.e., $d^{\Omega}_2=(a_\ast,0.5,5)$, etc. %The third individual is has zero productivity and twenty years of lifetime, i.e., $d^{\Omega}_3=(a_\ast,0,20)$. The fourth individual has $50\%$ productivity and five years of lifetime, i.e., $d^{\Omega}_4=(a_\ast,0.5,5)$. The last individual has zero productivity and lifetime, i.e., $d^{\Omega}_5=(a_\ast,0,0)$.
    
    In the second distribution ($d^{\Phi}$), the first individual has full health and maximal productivity as well as forty years of lifetime, i.e., $d^{\Phi}_1=d^{\Omega}_1=(a_\ast,1,40)$. The remaining individuals are experiencing health state $a$.\footnote{Capturing, for example, some problems in mobility, self-care, usual activities, pain/discomfort or anxiety/depression to consider the standard dimensions in the EuroQol system (e.g., Rabin and Charro, 2001).} They have zero productivity, except for the third individual, who has productivity $p$.\footnote{Reflecting, for instance, some level of absenteeism, presenteeism, or any other similar metric.} %The second individual has zero productivity and ten years of lifetime, i.e., $d^{\Phi}_2=(a,0,10)$, etc. %The third individual has productivity $p\in(0,1)$ and also ten years of lifetime, i.e., $d^{\Phi}_3=(a,p,10)$. The fourth individual has zero productivity and ten years of lifetime, i.e., $d^{\Phi}_4=(a,0,10)$. The last individual is experiencing zero productivity and lifetime, i.e., $d^{\Phi}_5=(a,0,0)$.
    
    \end{example}
    %\medskip

\subsection{Core axioms}

In what follows, we present some structural axioms for social preferences that are minimal and common to all evaluation functions in our framework.\footnote{The basic structural axioms were introduced in Hansen et al. (2024). They are adaptations of axioms in Østerdal (2005), Hougaard et al. (2013), and Moreno-Ternero et al. (2023) to the present model, which is enriched to account for productivity. Formal definitions are provided in the Appendix.}

%\vskip3mm\noindent \textbf{ANON}: $(d_1,...,d_n)\sim (d_{\pi(1)},...,d_{\pi(n)})$ for each $d\in D$, and each bijection $\pi: N\to N$.
%\medskip

%\vskip3mm\noindent \textbf{SEP}: For each pair $d,d'\in D$, and each $S\subset N$, $(d_S,d_{N\setminus S}) \succsim (d'_S,d_{N\setminus S}) \Leftrightarrow (d_S,d'_{N\setminus S}) \succsim (d'_S,d'_{N\setminus S})$. 
%\medskip

%\vskip3mm\noindent \textbf{CONT}: Let $d,d'\in D$, and $d^{(k)}$ be a sequence such that for each $i \in N$, $d_i^{(k)}=(a_i,p_i^{(k)}, t_i^{(k)}) \rightarrow (a_i,p_i, t_i)=d_i$. If $d_i^{(k)} \succsim d'$ for each $k$ then $d \succsim d'$, and if $d' \succsim d_i^{(k)}$ for each $k$, then $d' \succsim d$. 
%\medskip

%\vskip3mm\noindent \textbf{ZERO}: For each $d \in D$ and each $i \in N$ such that $t_i = 0$, and each $a^{\prime}_{i}\in A$, and each $p_i' \in [0,1]$, $d\sim [d_{N\setminus \{i\}}, (a_i', p_i', 0)]$. 
%\medskip

%\vskip3mm\noindent \textbf{FHPS}: For each $d\in D$ and each $i\in N$, $[(a_\ast,p_i, t_i),d_{N\setminus \{i\}}] \succsim d$ and $[(a_i,1, t_i),d_{N\setminus \{i\}}] \succsim d$.
%\medskip

%\vskip3mm\noindent \textbf{LMFHP}: For each $d \in D$ and each $i \in N$, such that $(a_i,p_i)=(a_\ast,1)$ and each $t_i > t_i'$, $[(a_\ast,1, t_i), d_{ N\setminus \{i\}}] \succ [(a_\ast,1,t_i'), d_{ N\setminus \{i\}}]$. 
%\medskip

%\vskip3mm\noindent \textbf{PLD}: For each $d \in D$ and each $i \in N$, $ d \succsim [(a_i,p_i,0), d_{N\setminus \{i\}}]$. 
%\medskip

The first axiom, \textit{anonymity}, states that it does not matter who holds which profile (only the distribution as a whole matters). In other words, swapping individuals' positions does not affect the evaluation.
%permuting triples does not alter preferences. 
The second axiom, \textit{separability}, says that if the profiles change only for a subgroup of individuals, then the relative evaluation of the two distributions should only depend on this subgroup.\footnote{In Example 1, the axiom implies that replacing individual 1 in both distributions by an individual with another profile would not alter the preferences between both distributions.} %Bleichrodt and Pinto (2005) consider a different notion of separability.
The third axiom, \textit{continuity}, says that small changes in productivity or life years should only produce small changes in the evaluation of the distribution. The \textit{social zero condition} says that if an individual has zero lifetime, then the health state and productivity of that individual do not influence the evaluation of the distribution.\footnote{In Example 1, this implies that changing the health state or productivity of individual 5 in each distribution would render the resulting distributions equally valued. Bleichrodt et al. (1997) and Miyamoto et al. (1998) introduce an analogous condition for individual preferences over health states.} 
The next axiom, \textit{full health and productivity superiority}, says that replacing an individual's health state by that of full health, or increasing productivity to its maximum, cannot worsen the evaluation of the distribution.\footnote{In Example 1, this implies that changing individual 3's health state (productivity) to full health (maximal productivity) in the second distribution does not worsen such a distribution.} 
Another natural axiom, \textit{lifetime monotonicity at full health and productivity}, says that if an individual enjoys full health and maximal productivity, then increasing life years is strictly better.\footnote{One might argue that weakening this axiom allows one to account for the existence of a satiation point or plateau. But our analysis focuses on realistic scenarios well below such threshold, for which increases in lifetime are socially desirable, as long as the individual is both at full health and is maximally productive.} The last structural axiom, \textit{positive lifetime desirability}, says that the distribution does not worsen if any individual moves from zero lifetime to positive lifetime when the health state and productivity are kept fixed. %This implies, in particular, that health states worse than death are excluded. 

We introduce an additional axiom, \textit{Pigou-Dalton transfer at full health and productivity} (PDTFHP), which will be central to our analysis, as it reflects a natural equity concern among fully healthy and productive individuals.\footnote{The counterpart axiom without considering productivity was introduced by Hougaard et al. (2013) and was also used by Moreno-Ternero and \O sterdal (2017). The notion of Pigou-Dalton transfers has a long tradition in inequality measurement (e.g., Cowell, 2000; Bosmans et al., 2009).}  
This axiom states that, all else equal, if we have two such individuals, then a hypothetical transfer of lifetime from the one with more to the one with less leads to a more desirable distribution, as long as the transfer is not so large that it reverses their relative positions. 

In what follows, we shall refer to the whole pack of axioms introduced above as the \textit{CORE} axioms.\footnote{CORE without PDTFHP is what Hansen et al. (2024) call COMMON.}

\subsection{Baseline structures}

Each individual profile $d_i=(a_i, p_i, t_i)$ can be associated with another profile $(a_\ast,1, f(a_i,p_i, t_i))$ in which the individual has full health and maximal productivity, but a lower lifetime. The lifetime amount, $f(a_i,p_i, t_i)$, which might depend on the whole initial triple, is called the \textit{healthy productive years equivalent} (\textit{HPYE}) for individual $i$, and is determined such that the policymaker is indifferent between the original distribution and the new one. Formally,
let $ f :A\times [0,1]\times \Bbb{R}_+\rightarrow \mathbb{R}_+$ be a function, which is continuous with respect to its second and third variables, where for each $d=[d_1,\dots,d_n]=[(a_1,p_1,t_1),\dots, (a_n,p_n,t_n)] \in D$, 
$$d \sim [(a_\ast,1, f(a_i,p_i, t_i))_{i\in N}],$$
and, for each $(a_i,p_i,t_i) \in A\times [0,1]\times \Bbb{R}_+$, $0 \le f(a_i,p_i,t_i) \le t_i$,  $f(a_i,p_i,t_i) \le f(a_i,1,t_i)$, and $f(a_i,p_i,t_i) \le f(a_\ast,p_i,t_i)$. In particular, $f(a_*,1, t_i)=t_i$.

As shown in Hansen et al. (2024), the structural axioms above ensure that $f$ is well-defined and uniquely determined.\footnote{In particular, lifetime monotonicity at full health and productivity is what guarantees uniqueness.} 

As mentioned in the introduction, healthy productive years equivalent can be seen as an extension of healthy-years equivalents, introduced by Mehrez and Gafni (1989), to account for productivity. A somewhat similar concept that has also been used in health economics is \textit{equivalent income} (e.g., Fleurbaey et al., 2013).\footnote{If %each individual is depicted by $(y_i,h_i,z_i)$, rather than $(a_i,p_i,t_i)$, where 
$y_i$ denotes $i$'s income, $h_i$ denotes $i$'s health, and $z_i$ denotes other dimensions of well-being (e.g., leisure, public goods, etc.), $i$'s equivalent income $y^*_i$ is a level of income such that $i$ is indifferent between $(y_i,h_i,z_i)$ and $(y^*_i,h^*,z^*)$, where $h^*$ denotes full health and $z^*$ denotes a reference value of $z$. Note that equivalent income is defined for individual preferences, whereas healthy productive years equivalent is defined for the preferences of a policymaker.} 

The so-called \textit{generalized HPYE} evaluation function is defined by the aggregation of some function of the individual HPYEs.  
Formally,
\begin{equation}\label{eqgenHPYE}
 E^g[d_1,\dots,d_n]= E^g[(a_1,p_1,t_1),\dots, (a_n,p_n,t_n)]=\sum_{i=1}^n g(f(a_i,p_i,t_i)),
\end{equation}
where $g: \mathbb{R}_{+} \rightarrow \mathbb{R}$ is a strictly increasing and continuous function.

Of particular interest are those %generalized HPYE evaluation functions 
arising when $g$ is a concave function. The following result states that such a class of equity-sensitive generalized HPYE evaluation functions %from Hansen et al. (2024) 
is precisely characterized by the set of CORE axioms. 
%, together with the Pigou-Dalton transfer at full health and productivity. 
This result will be the foundation for alternative models we present in this paper. 
\begin{theorem}\label{thm:genHPYE}
    The following statements are equivalent:
    \begin{enumerate}
     \item $\succsim $ is represented by a generalized HPYE evaluation function (\ref{eqgenHPYE}) with a strictly concave $g$ function.
    \item $\succsim $ satisfies CORE. %and PDTFHP.
    \end{enumerate} 
\end{theorem}
\medskip

Note that Theorem \ref{thm:genHPYE} excludes the special case of (\ref{eqgenHPYE}) where $g$ is the identity function. That is, the (unweighted) \textit{HPYE} evaluation function, which assesses distributions by aggregating the HPYEs they yield. Formally,
\begin{equation} \label{eq:*}\tag{*}
E^f[d_1,\dots,d_n]= E^f[(a_1,p_1,t_1),\dots, (a_n,p_n,t_n)]=\sum_{i=1}^n f(a_i,p_i,t_i).
\end{equation}  
The evaluation function (\ref{eq:*}) is utilitarian in spirit and therefore does not address distributional concerns. While we do not characterize it here, we
present it for later reference and comparisons.\footnote{It is characterized in Hansen et al. (2024).}

%\medskip 

\section{General power evaluation functions} 
The generalized HPYE evaluation function described above is highly flexible and can be adapted to a wide range of applications. For practical purposes, however, a one-parameter specification of $g$ offers substantial analytical and computational convenience. For this reason, we focus on a specific form where $g$ is a concave power function. This specification not only simplifies implementation but also captures a central equity concern in lifetime evaluation: diminishing marginal value of additional years (at full health and maximal productivity). A concave power function reflects the intuitive and normative idea that extending a shorter life should carry greater weight than extending an already long life, thus aligning with widely accepted principles of distributive justice.
Formally, 
\begin{equation}\label{gammaf}
    E^{\gamma f}[d_1,\dots,d_n]= E^{\gamma f}[(a_1,p_1,t_1),\dots, (a_n,p_n,t_n)]=\sum_{i=1}^n f(a_i,p_i,t_i)^{\gamma},
    \end{equation}
    where $0<\gamma<1$ and $f$ is the HPYE function.  

\medskip

The parameter $\gamma$ captures the degree of equity sensitivity: lower values of $\gamma$ reflect stronger inequality aversion by placing relatively greater weight on gains to individuals with shorter healthy productive lifetimes. Conversely, higher values of $\gamma$ reduce this concern, approaching a utilitarian aggregation of HPYEs. It turns out that this family is characterized when adding a new axiom, \textit{time scale independence at full health and productivity}, to the pack of CORE axioms. This axiom says that, if everyone has full health and maximal productivity, the ranking of a pair of distributions does not reverse when all lifetimes are multiplied by a common positive constant. 
%It reflects the normative view that only relative lifetime length is ethically relevant, and that the specific temporal units used to measure lifetimes (e.g., years versus months) carry no intrinsic ethical significance, when everyone has full health and maximal productivity, distributions can be evaluated independently of the specific units used to measure lifetimes (e.g., days, months, or years).
In such cases, only relative lifetime is normatively relevant, and the units used to measure lifetimes carry no intrinsic significance.\footnote{For a treatment of scale independence in a welfarist context, see, for instance, Moulin (1988) and Morreau and Weymark (2016).} Formally,

\medskip 

\vskip3mm\noindent \textbf{TSIFHP}: For each pair $d=[(a_{\ast},1,t_1),\dots, (a_{\ast},1,t_n)]\in D,d^{\prime}=[(a_{\ast}, 1, t'_1),\dots, (a_{\ast},1,t'_n) ]\in D$, and each $c>0$, 
$$d\succsim d^{\prime} \iff [(a_{\ast},1,ct_1),\dots, (a_{\ast},1,ct_n)]\succsim [(a_{\ast}, 1, ct'_1),\dots, (a_{\ast},1,ct'_n)].$$ 

%\medskip

\begin{theorem}\label{thm:powerf}
    The following statements are equivalent:
    \begin{enumerate}
     \item $\succsim $ is represented by an equity-sensitive power HPYE evaluation function (\ref{gammaf}).
    \item $\succsim $ satisfies CORE and TSIFHP.
    \end{enumerate} 
\end{theorem}

%The power family of HPYE representations implied by time scale independence 
The family $E^{\gamma f}$ %(\ref{gammaf}) 
just characterized exhibits strong priority to individuals with very short lifetimes. More precisely, for two individuals $(a_{\ast}, 1, t_1)$ and $(a_{\ast}, 1, t_2)$, %at full health and productivity, equation  implies
$$\frac{\frac{\partial E^{\gamma f}}{\partial t_1}}{\frac{\partial E^{\gamma f}}{\partial t_2}}=\left(\frac{t_2}{t_1}\right)^{1-\gamma},$$
which increases without bound as $t_1$ approaches zero (recall that $0<\gamma<1$). This reflects the fact that marginal social value is higher for lifetime gains accruing to those who are worst off in terms of time.
Importantly, this priority applies to marginal, infinitesimal lifetime changes. The social value of any finite increase in lifetime remains bounded. That is, for each $\Delta>0$, $(t+\Delta)^\gamma-t^\gamma\to \Delta^\gamma$ as $t\to 0$. The representation therefore combines strong concern for individuals with extremely short lifetimes with finite social valuation of concrete policy-relevant lifetime gains. In this sense, time scale independence yields a particularly sharp formalization of prioritarian concern (e.g., Adler, 2012) in lifetime allocation.
\medskip

A natural generalization of TSIFHP arises when the notion applies to arbitrary levels of health and productivity. We simply refer to the resulting axiom as \textit{time scale independence}.\footnote{The axiom was also considered by Østerdal (2005) and Hougaard et al. (2013) in the context of social preferences regarding health distributions, excluding productivity considerations. The individual-level counterpart of TSI is known in the literature as the \textit{constant proportional trade-off assumption} (e.g., Pliskin et al., 1980).}
\medskip

\vskip3mm\noindent \textbf{TSI}: For each pair $d=[(a_1,p_1,t_1),\dots, (a_n,p_n,t_n)]\in D,d^{\prime}=[(a'_1, p'_1, t'_1),\dots, (a'_n,p'_n,t'_n) ]\in D$, and each $c>0$,
$$d\succsim d^{\prime} \iff [(a_1,p_1,ct_1),\dots, (a_n,p_n,ct_n)]\succsim [(a_1', p_1', ct'_1),\dots, (a'_n,p'_n,ct'_n)].$$ 

\medskip

When this axiom replaces TSIFHP at Theorem \ref{thm:powerf}, we characterize an evaluation function in which time enters via a (concave) power function that is adjusted to both quality of health and productivity. We will refer to it as an \textit{equity-sensitive productivity- and quality-adjusted power lifetime} evaluation function. Formally, %, i.e., $ \varphi(a_i,p_i)=q(a_i)p_i^{\gamma}$, for each $(a_i,p_i)\in A\times[0,1]$, where $\gamma\in (0,1)$. 
\begin{equation}\label{gamma}
    E^{\gamma\varphi}[d_1,\dots,d_n]= E^{\gamma\varphi}[(a_1,p_1,t_1),\dots, (a_n,p_n,t_n)]=\sum_{i=1}^n \varphi(a_i,p_i)t_i^{\gamma},
    \end{equation}  
    where $\varphi:A\times [0,1] \rightarrow [0,1]$ is a continuous function in its second variable such that $0\le \varphi(a,p)\le \varphi(a_{\ast},1)=1$, $\varphi(a,p) \le \varphi(a_{\ast},p),$ and $\varphi(a,p) \le \varphi(a,1)$ for each $(a,p)\in A\times[0,1]$, and $0<\gamma <1$. 
\medskip

\begin{theorem}\label{thm:power}
    The following statements are equivalent:
    \begin{enumerate}
     \item $\succsim $ is represented by an equity-sensitive productivity- and quality-adjusted power lifetime evaluation function (\ref{gamma}).
    \item $\succsim $ satisfies CORE and TSI.
    \end{enumerate} 
\end{theorem}

Two extreme members of the previous family are obtained when the adjustment only depends on quality or productivity (but not both). This gives rise to %egalitarian quality-productivity-adjusted power lifetime evaluation function. 
the basic \textit{equity-sensitive power QALY} and \textit{equity-sensitive power PALY} evaluation functions. 
Formally,
\begin{equation}  \label{power-eqQALY}
E^{\gamma q}[d_1,\dots,d_n]= E^{\gamma q}[(a_1,p_1,t_1),\dots, (a_n,p_n,t_n)]=\sum_{i=1}^n q(a_i)t^{\gamma}_i,
\end{equation}
where $q:A\to[0,1]$ is a function satisfying $0 \le q(a_{i})\le
q(a_{\ast})=1$, for each $a_{i}\in A$, and $0<\gamma<1$.
\begin{equation}\label{power-eqPALY}
    E^{\gamma p}[d_1,\dots,d_n]= E^{\gamma p}[(a_1,p_1,t_1),\dots, (a_n,p_n,t_n)]=\sum_{i=1}^n p_it^{\gamma}_i,
   \end{equation}
where $0<\gamma<1$.

The previous evaluation functions ignore, respectively, productivity and health. More precisely, they respectively satisfy the axioms of \textit{productivity independence} and \textit{health independence} introduced next. The first of these axioms specifies that if the health state and lifetime of an individual are fixed, then evaluation of the distribution does not depend on the productivity level. Similarly, the second axiom says that if the productivity level and lifetime are fixed, then the health state of an individual will not affect the evaluation.\footnote{Note that both axioms should not be read as expressing separability across attributes, but rather as stating that the evaluation does not depend on the attribute in question.}

\vskip3mm\noindent \textbf{PI}: For each $d \in D$, each $i \in N$ and $p_i'\not =p_i$, $ [(a_i,p_i,t_i), d_{N\setminus\{i\}}] \sim [(a_i,p_i',t_i), d_{N\setminus\{i\}}]$. 
\bigskip

\vskip3mm\noindent \textbf{HI}: For each $d \in D$, each $i \in N$ and $a_i'\not =a_i$, $[(a_i,p_i,t_i), d_{ N\setminus \{i\}}] \sim [(a'_i,p_i,t_i), d_{ N\setminus \{i\}}]$.
\bigskip

The following result emerges from Theorem \ref{thm:power}. It also constitutes the extension to our setting of the counterpart result in Hougaard et al. (2013).

\begin{proposition}\label{power-QALY}
The following statements are equivalent:
\begin{enumerate}
\item $\succsim $ is represented by an equity-sensitive power QALY evaluation function (\ref{power-eqQALY}). %PHEFPHEF %evaluation% satisfying $(\ref{QU})$.
\item $\succsim $ satisfies CORE, PI, and TSI. 
\end{enumerate}
\end{proposition}

The counterpart result for equity-sensitive power PALYs requires two extra axioms to be considered next.\footnote{These two axioms were considered in Hansen et al. (2024). Note that a weaker version of the first one, restricting only to cases of full health would suffice for the characterization.}

First, \textit{productivity invariance at common health and time}, which says that for two individuals at common health and time, it does not matter who gains in productivity.

\vskip3mm\noindent \textbf{PICHT}: For each $d\in D$, each pair $i,j\in N$ with $a_i=a_j=a$ and $t_i=t_j=t$, and each $c>0$ such that $p_i+c,p_j+c \leq 1$,
$$[(a,p_i+c,t),(a,p_j,t), d_{ N\setminus \{i,j\}}] \sim [(a,p_i,t),(a,p_j+c,t), d_{ N\setminus \{i,j\}}].$$

Second, \textit{time independence for unproductive individuals}, which states that if an individual has zero productivity, then the lifetime of that individual does not influence the evaluation of the distribution.%irrelevance versus independence. Can we find something weaker that does the job?

\vskip3mm\noindent \textbf{TIUP}: For each $d \in D$ and each $i \in N$ such that $p_i = 0$, and each $t^{\prime}_{i}\in \Bbb{R}_+$, $d\sim [d_{N\setminus \{i\}}, (a_i, 0, t^{\prime}_{i})]$. 

\begin{proposition}\label{power-PALY} 
    The following statements are equivalent:
\begin{enumerate}
\item $\succsim $ is represented by an equity-sensitive power PALY evaluation function (\ref{power-eqPALY}).
\item $\succsim $ satisfies CORE, HI, TSI, PICHT, and TIUP. 
\end{enumerate}
\end{proposition}

Health and productivity both influence individuals' equivalent lifetimes in the HPYE framework, but they play different roles in the axiomatic analysis. The set of health states is left without intrinsic mathematical structure and affects social evaluation only through equivalence to lifetime at the benchmark of full health and productivity. In contrast, productivity is measured on a bounded scale and is subject to axioms governing how lifetime is evaluated for individuals with zero productivity. As no health state is treated as rendering lifetime socially irrelevant, there is no corresponding health analogue of these productivity-based restrictions. The axiomatic characterizations presented above therefore reflect distinct normative assumptions applied to different attributes within a common HPYE structure.\footnote{One could nevertheless obtain a counterpart of Proposition \ref{power-QALY} with the axioms CORE, HI, and TSI, characterizing an equity-sensitive \textit{generalized power PALY} evaluation function in which productivity enters via a continuous function $v:[0,1] \to [0,1]$ such that $v(1)=1$.}
%via a function from its domain onto itself.}

We conclude this section by applying the last two evaluation functions to the distributions from Example 1. 

To illustrate, we assume that $a$ is the health state 11211 in the EQ-5D-5L system (that is, a state with slight problems on usual activities and no problems in mobility, self-care, pain/discomfort, and anxiety/depression). According to Devlin et al. (2018), the value associated to $a$ in England is $0.95$, i.e., we assume $q(a)=0.95$ in our analysis. We also assume $p=\frac{1}{\sqrt{2}}\approx 0.71$. Then, we obtain the following:
$$E^{\gamma q}[d^{\Omega}]\ge E^{\gamma q}[d^{\Phi}] \text{ if and only if }\gamma\in\left(0,0.32\right]\cup \left[0.68, 1\right),  \text{ and }$$ 
$$E^{\gamma p}[d^{\Omega}]\ge E^{\gamma p}[d^{\Phi}] \text{ if and only if }\gamma\in\left(0,0.5\right].$$ 

This result highlights that the preference between the two distributions may depend non-monotonically on the equity-sensitivity parameter~$\gamma$. For the quality-based evaluation, this is because low values of~$\gamma$ place high weight on short lifespans, favoring $\Omega$ due to its severely short-lived individuals, while high values emphasize total health-adjusted time, again favoring $\Omega$. At intermediate~$\gamma$, the more balanced distribution in~$\Phi$ may be preferred.
In contrast, the productivity-based evaluation favors~$\Omega$ only when $\gamma \le \tfrac{1}{2}$, as sufficiently little weight is then placed on longer lives. For higher values of $\gamma$, the longer productive life in~$\Phi$ can dominate, despite lower productivity. 

\section{Equity-sensitive compromises of QALYs and PALYs}\label{Power PQALYs}

In this section, we explore more specialized evaluation functions, which compromise between QALYs and PALYs and may be particularly useful for practical applications. 

The next evaluation function, which we dub the \textit{equity-sensitive power PQALY} 
evaluation function, is a focal equity-sensitive compromise between QALYs and PALYs. It evaluates distributions by aggregating power-transformed lifetimes, weighted according to individuals' health and productivity levels, such that health, productivity, and lifetime duration enter the function multiplicatively.  
Formally,
\begin{equation}\label{power-eqM-PQALY}
E^{\gamma pq}[d_1,\dots,d_n]= E^{\gamma pq}[(a_1,p_1,t_1),\dots, (a_n,p_n,t_n)]=\sum_{i=1}^n q(a_i)p_it^{\gamma}_i,
\end{equation}
where $q:A\rightarrow [0,1]$ is a function satisfying $0\le q(a_i) \le q(a_\ast)= 1$ 
for each $a_i\in A$, and $0<\gamma<1$.

As the next result states, this evaluation function is characterized when we dismiss the health independence axiom used in Proposition \ref{power-PALY}. 

\begin{theorem}\label{thm:power-M-PQALY} 
The following statements are equivalent:
\begin{enumerate}
\item $\succsim $ is represented by an equity-sensitive power PQALY evaluation function (\ref{power-eqM-PQALY}).
\item $\succsim $ satisfies CORE, TSI, PICHT, and TIUP. 
\end{enumerate}
\end{theorem}

The next family of evaluation functions, which we dub the \textit{equity-sensitive power QALY-PALY} evaluation functions, evaluate distributions by taking convex combinations of the equity-sensitive power QALYs and the equity-sensitive power PALYs the distribution yields. Formally,
\begin{equation}\label{power-QALY-PALY}
E^{\gamma\sigma}[d_1,\dots,d_n]= E^{\gamma\sigma}[(a_1,p_1,t_1),\dots, (a_n,p_n,t_n)]=\sigma\sum_{i=1}^n q(a_i)t^{\gamma}_i+(1-\sigma)\sum_{i=1}^n p_it^{\gamma}_i,
\end{equation}
where $q:A\rightarrow [0,1]$ is a function satisfying $0\le q(a_i) \le q(a_\ast)= 1$, 
for each $a_i\in A$, $\sigma\in[0,1]$, and $0<\gamma<1$.

This family of evaluation functions is characterized when the TIUP axiom is dismissed from the statement of Theorem \ref{thm:power-M-PQALY} and when the PICHT axiom is strengthened to \textit{productivity invariance at common time}, 
%is not imposed, and when the \textit{productivity invariance at common health and time} is strengthened to \textit{productivity invariance at common time}, 
which says that for two individuals with common lifespan (and irrespective of their health status), it does not matter which of the two individuals improves their productivity. Formally,

\vskip3mm\noindent \textbf{PICT}: For each $d\in D$, each pair $i,j\in N$ with $t_i=t_j=t$, and each $c>0$ such that $p_i+c,p_j+c \leq 1$,
$$[(a_i,p_i+c,t),(a_j,p_j,t), d_{ N\setminus \{i,j\}}] \sim [(a_i,p_i,t),(a_j,p_j+c,t), d_{ N\setminus \{i,j\}}].$$

\begin{theorem}\label{thm:power-QALY-PALY} 
The following statements are equivalent:
\begin{enumerate}
 \item $\succsim $ is represented by an equity-sensitive power QALY-PALY evaluation function (\ref{power-QALY-PALY}).
\item $\succsim $ satisfies CORE, TSI and PICT. 
\end{enumerate}
\end{theorem}

%Based on the last two results, one might wonder what lies between the ethics of \textit{productivity invariance at common health and time} and \textit{productivity invariance at common time}, without resorting to \textit{time independence for unproductive individuals}. For instance, how should we treat (from a social viewpoint) extra productivity, when accruing to an individual enjoying full health or to another individual not enjoying full health? The \textit{productivity monotonicity at common time} axiom states the former should be preferred to the latter. Formally, 

%\vskip3mm\noindent \textbf{PMCT}: For each $d\in D$, each pair $i,j\in N$ with $t_i=t_j=t$, and each $c>0$ such that $p_i+c,p_j+c \leq 1$,
%$$[(a_{\ast},p_i+c,t),(a_j,p_j,t), d_{ N\setminus \{i,j\}}] \succsim [(a_{\ast},p_i,t),(a_j,p_j+c,t), d_{ N\setminus \{i,j\}}].$$

Based on the last two results, one might wonder what is obtained when replacing PICT with PICHT in the previous statement, or, equivalently, by omitting the TIUP axiom at the statement of Theorem \ref{thm:power-M-PQALY}. It turns out that we arrive at an alternative and intriguing compromise, which we refer to as the \textit{equity-sensitive power QALY-PQALY} evaluation. This function assesses distributions via a combination of equity-sensitive power QALYs and equity-sensitive power PQALYs, with weights split between the two components. 
%generalized combination of the equity-sensitive power QALYs and the equity-sensitive power PQALYs generated by the distribution. 
Formally,
\begin{equation}\label{power-eqA-PQALY}
E^{\gamma rs}[d_1,\dots,d_n]= E^{\gamma rs}[(a_1,p_1,t_1),\dots, (a_n,p_n,t_n)]=\sum_{i=1}^n r(a_i)t^{\gamma}_i+\sum_{i=1}^n s(a_i)p_it^{\gamma}_i,
\end{equation}
where $r,s:A\rightarrow [0,1]$ are functions satisfying $0\le r(a_i) \le r(a_\ast)\le 1$, $0\le s(a_i)\le 1$, and $0\le r(a_i) +s(a_i)\le r(a_\ast)+s(a_\ast)= 1$ for each $a_i\in A$, and $0<\gamma<1$.\footnote{The functional form (\ref{power-eqA-PQALY}) includes as a focal element the family made of the convex combinations between the equity-sensitive power QALYs (\ref{power-eqQALY}) and the equity-sensitive power PQALYs (\ref{power-eqM-PQALY}). Such a family could be characterized by adding to the statement of Theorem \ref{thm:power-A-PQALY} a natural assumption stating that, from a social viewpoint, extra productivity accruing to an individual at full health is at least as good as when it accrues to an individual at another health state.}

%Without the PMCT axiom (and the TIUP axiom) we would characterize. More precisely, 

\begin{theorem}\label{thm:power-A-PQALY} 
The following statements are equivalent:
    \begin{enumerate}
    \item $\succsim $ is represented by an equity-sensitive power QALY-PQALY evaluation function (\ref{power-eqA-PQALY}).
    \item $\succsim $ satisfies CORE, TSI and PICHT. 
    \end{enumerate}
\end{theorem}

The last evaluation function within this section, which we dub the \textit{equity-sensitive bi-power PQALY} %productivity-and-quality-adjusted life years} (power PQALY)
evaluation function, is another focal equity-sensitive compromise between QALYs and PALYs. It evaluates distributions by aggregating power-transformed lifetimes, weighted according to individuals' health and power-transformed productivity levels, such that health, productivity, and lifetime factors enter the function multiplicatively.  
Formally, 
\begin{equation}\label{gammam-core}
        E^{\gamma \varepsilon}[d_1,\dots,d_n]= E^{\gamma \varepsilon}[(a_1,p_1,t_1),\dots, (a_n,p_n,t_n)]=\sum_{i=1}^n q(a_i)p^\varepsilon_i t_i^{\gamma},
        \end{equation}  
    where $q:A\to[0,1]$ is a function satisfying $0 \le q(a_{i})\le q(a_{\ast})=1$, for each $a_{i}\in A$, and $0<\varepsilon,\gamma<1$. 
\medskip

This bi-power representation separates the treatment of lifetime and productivity. The parameter $\gamma$ governs inequality aversion in lifetime, while the parameter $\varepsilon$ governs inequality aversion in	 productivity. For example, choosing a small value of $\varepsilon$ limits the extent to which productivity disparities influence the social evaluation, enabling a policymaker to moderate the weight assigned to productivity.

To capture how productivity should be treated in this setting, a natural counterpart of the TSI axiom for productivity is \textit{productivity scale independence}, which says that social preferences are invariant to common rescaling of productivity. Formally, 

\medskip

\vskip3mm\noindent \textbf{PSI}: For each pair $d=[(a_1,p_1,t_1),\dots, (a_n,p_n,t_n)]\in D,d^{\prime}=[(a'_1, p'_1, t'_1),\dots, (a'_n,p'_n,t'_n) ]\in D$, and each $0<c<1$,
$$d\succsim d^{\prime} \iff [(a_1,cp_1,t_1),\dots, (a_n,cp_n,t_n)]\succsim [(a_1', cp_1', t'_1),\dots, (a'_n,cp'_n,t'_n)].$$ 

\medskip

Likewise, a natural counterpart of the PDTFHP axiom for productivity is \textit{Pigou-Dalton transfer at full health and common time}, which says that if we have two individuals at full health and common lifetime, \textit{transferring productivity} from the one with more to the one with less, ceteris paribus, results in a better distribution. Formally,

\vskip3mm\noindent \textbf{PDTFHCT}: For each $d=[(a_{\ast},p_k,t_{k})_{k\in N}]$, and each pair $i,j\in N$, such that $t_{i}= t_{j}$, $p_{i}< p_{j}$ and each $0<\rho<\frac{p_{j}-p_{i}}{2}$, 
\begin{equation*}
\left[ \left(a_{\ast},p_{i}+\rho,t_i\right), \left(a_{\ast},p_{j}-\rho,t_{j}\right),d_{N\backslash \{i,j\}}\right]\succ d.
\end{equation*}

\medskip

As stated in the next result, the equity-sensitive bi-power PQALY evaluation function is characterized when we add the last two axioms to TSI to CORE. 

\begin{theorem}\label{thm:bi-power-PQALY} 
The following statements are equivalent:
\begin{enumerate}
\item $\succsim $ is represented by an equity-sensitive bi-power PQALY evaluation function (\ref{gammam-core}).
\item $\succsim $ satisfies CORE, TSI, PSI and PDTFHCT. 
\end{enumerate}
\end{theorem}

The next table summarizes our results.

\bigskip

\bigskip

%\begin{equation*}
\begin{tabular}{|c|c|c|c|c|c|c|c|c|c|}
\hline
\begin{tabular}{ll}
& $E$ \\ 
Axioms & 
\end{tabular}
& \begin{tabular}{l} $E^g$ \\ (\ref{eqgenHPYE})\end{tabular}   
& \begin{tabular}{l} $E^{\gamma f}$ \\ (\ref{gammaf})\end{tabular}
& \begin{tabular}{l} $E^{\gamma \varphi}$ \\ (\ref{gamma})\end{tabular} 
& \begin{tabular}{l} $E^{\gamma q}$ \\ (\ref{power-eqQALY})\end{tabular} 
& \begin{tabular}{l} $E^{\gamma p}$ \\ (\ref{power-eqPALY})\end{tabular}
& \begin{tabular}{l} $E^{\gamma pq}$ \\ (\ref{power-eqM-PQALY})\end{tabular}
& \begin{tabular}{l} $E^{\gamma \sigma}$ \\ (\ref{power-QALY-PALY})\end{tabular}
& \begin{tabular}{l} $E^{\gamma rs}$ \\ (\ref{power-eqA-PQALY})\end{tabular}
& \begin{tabular}{l} $E^{\gamma \varepsilon}$ \\ (\ref{gammam-core})\end{tabular}
\\ \hline
\multicolumn{1}{|l|}{CORE} & YES$^{\ast}$ & YES$^{\ast}$ & YES$^{\ast}$ & YES$^{\ast}$ & YES$^{\ast}$ & YES$^{\ast}$& YES$^{\ast}$& YES$^{\ast}$& YES$^{\ast}$ \\ \hline
\multicolumn{1}{|l|}{PI} & NO & NO & NO & YES$^{\ast}$ & NO& NO& NO& NO& NO \\ \hline
\multicolumn{1}{|l|}{HI} & NO & NO & NO & NO & YES$^{\ast}$& NO& NO& NO& NO \\ \hline
\multicolumn{1}{|l|}{TSIFHP} & NO & YES$^{\ast}$ & YES & YES & YES& YES& YES& YES& YES \\ \hline
\multicolumn{1}{|l|}{TSI} & NO & NO & YES$^{\ast}$ & YES$^{\ast}$ & YES$^{\ast}$ & YES$^{\ast}$ & YES$^{\ast}$ & YES$^{\ast}$ & YES$^{\ast}$  \\ \hline
\multicolumn{1}{|l|}{PICHT} & NO & NO & NO & YES & YES$^{\ast}$ & YES$^{\ast}$ &  YES& YES$^{\ast}$& NO\\ \hline
\multicolumn{1}{|l|}{PICT} & NO & NO & NO & YES & YES& NO& YES$^{\ast}$ & NO &  NO\\ \hline
%\multicolumn{1}{|l|}{PMCT} & NO & NO & NO & YES & NO& NO& NO & YES$^{\ast}$  &  NO\\ \hline
\multicolumn{1}{|l|}{TIUP} & NO & NO & NO & NO & YES$^{\ast}$ & YES$^{\ast}$& NO& NO& YES \\ \hline
\multicolumn{1}{|l|}{PSI} & NO& NO & NO & YES & YES & YES& NO& NO& YES$^{\ast}$\\ \hline
\multicolumn{1}{|l|}{PDTFHCT} & NO & NO & NO & NO & NO & NO& NO& NO& YES$^{\ast}$\\ \hline
%\multicolumn{1}{|l|}{PDTPH} & NO & NO & iff $0<\gamma < 1$& iff $0<\gamma < 1$ & iff $w$ strictly concave \\ \hline
\end{tabular}%
\ \ 

\bigskip

%\begin{equation*}
%\text{
\textbf{Table 1: Overview of evaluation functions and underlying axioms}. 
“YES” means that the corresponding evaluation function satisfies the given axiom, while “NO” means it does not. An asterisk next to “YES” indicates that the axiom is used to characterize the corresponding evaluation function.
%}
%\end{equation*}

\bigskip

We also conclude this section by applying the evaluation functions characterized above to the distributions from Example 1. In general, we find that the ranking between the two distributions depends on the parameter values defining each family.  
%In general, we obtain that a distribution is preferred to the other for a certain range of the parameters defining the families. %of evaluation functions we are considering in this section. 
As before, we take $a$ to be the health state 11211 in the EQ-5D-5L system, with $q(a)=0.95$. We now set $p=0.5=\gamma$. 
%and $q(a_{\ast})=1=r(a_{\ast})$. 
Moreover, for each $\delta\in (0,1)$, we let $r(a_{\ast})=\delta$, $s(a_{\ast})=1-\delta$, $r(a)=0.95\delta$, $s(a)=0.95(1-\delta)$. Then, we obtain the following:
$$E^{\gamma pq}[d^{\Omega}]\ge E^{\gamma pq}[d^{\Phi}],$$ %for each $\gamma\in [0,1]$.%and  
$$E^{\gamma \sigma}[d^{\Omega}]\ge E^{\gamma \sigma}[d^{\Phi}]\text{ if and only if } 0\le \sigma\le 0.90, $$  
$$E^{\gamma rs}[d^{\Omega}]\ge E^{\gamma rs}[d^{\Phi}]\text{ if and only if } 0 < \delta\le 0.91,\text{ and }$$ 
$$E^{\gamma \varepsilon}[d^{\Omega}]\ge E^{\gamma \varepsilon}[d^{\Phi}]\text{, for all } 0< \varepsilon< 1. $$ 

The examples above show that the relative evaluation of the two distributions depends on the specific functional form and parameter values used. In all cases, there exists a range of parameters for which $d^\Omega$ is preferred, but these ranges vary across the functions. This reflects how each function places different emphasis on health state, productivity, and time. Functions combining health state and productivity in a multiplicative way tend to favor $d^\Omega$ more robustly, while those that separate the two components or allow trade-offs through convex combinations show greater sensitivity to parameter changes. These patterns illustrate the importance of aligning the chosen evaluation method with the normative priorities in each context.

\section{Discussion} 
This paper has developed an integrated approach to outcome measurement that jointly considers health, productivity and equity concerns. The analysis bridges the conventional focus on quality-adjusted life years with the objective of enhancing productive capacity and reducing inequalities. In doing so, it contributes to the longstanding discussion on equitable health care resource allocation, as well as to the developing field of economic evaluation of OHS interventions (e.g., Culyer and Wagstaff, 1993; Tompa et al., 2008; Searcy et al., 2016; Steel et al., 2018). The framework generalizes traditional effectiveness criteria by allowing compromises between pure health gains and broader societal benefits, providing an equity-oriented normative basis for evaluating interventions that impact both health and productivity. Our main finding is that it is both feasible and conceptually meaningful to integrate equity and productivity considerations into health evaluation. This integration offers decision-makers a richer informational basis for health policy: one that acknowledges not only how much health is gained, but also who gains it and with what implications for economic productivity.
 
While the framework represents one viable approach, it also raises complex questions that warrant further investigation (as we outline below). 
We take it to represent a basis for advancing health valuation models that better reflect concerns of equity and productivity. Refining these models, addressing the value judgments they involve, and testing them in practical policy environments remain important tasks for researchers and policymakers alike. The potential payoff is a health evaluation paradigm that more accurately reflects who benefits and in what ways, supporting decisions that are both sustainable and equitable.

\subsection{Limitations} %Alongside its contributions, 
Our approach has several limitations that warrant careful consideration. 

First, the integration of equity concerns inevitably relies on normative judgments about the social value of health gains to different individuals or groups. We have adopted an axiomatic social welfare framework to justify equity-sensitive evaluation functions (e.g., power QALY models), but the choice of specific functions or parameters can be debated. Different societies may disagree on the degree to which an extra QALY for a disadvantaged person should count more than one for someone already healthy (e.g., Bleichrodt, 1997; Dolan, 1998). In practice, eliciting consensus on these trade-offs is challenging.\footnote{Empirical studies using discrete choice experiments find that public preferences for distributional weights can be heterogeneous (e.g., Clark et al., 2014).} Related to this, our results have excluded $\gamma<0$ in our equity-sensitive evaluation functions due to the possibility of zero lifetime values. Alternatively, one could exclude zero lifetimes from our domain and work only with strictly positive lifetimes (with the proviso that quality of life improvements become almost insignificant when lifetimes are small enough to become negligible).\footnote{This approach was taken by Moreno-Ternero and \O sterdal, (2017) for a model in which productivity was absent.} The axiomatic analysis in this paper could then be extended to characterize other functional forms, such as logarithmic or additively separable with negative powers (i.e., those excluded in the proofs of Theorems 2 and 3). Those functional forms would allow us to get closer to Robson et al. (2024), who elicit preferences by asking participants to allocate resources while revealing the consequences for the distribution of health.\footnote{Their model is nevertheless different, as the input is a distribution of QALYs and a (utilitarian) social welfare function is considered via an iso-elastic function. That is, $U(h_i) = \frac{h_i^{1-\varepsilon}}{1-\varepsilon}$, for $\varepsilon \ge 0$ and $\varepsilon \ne 1$, and $U(h_i)=\ln (h_i)$ for $\varepsilon = 1$. %Each individual $i$ is endowed with an amount $h_i\ge 0$, which is generated by the (linear) health production function, $h_i=p_iy_i$, where $y_i$ is the amount of resources $i$ gets in the allocation. Robson et al. (2024) note that a concave production function would be more realistic but would complicate the elicitation task. 
They estimate a median value of $\varepsilon=3.170$, with 
$0<\varepsilon<1$ for less than $19\%$. } %, i.e., .. .
 %They estimate the median is $\varepsilon=3.170$, with $0<\varepsilon<1$ for less than $19\%$. Now, in our case, (as mentioned above) the input is a distribution of triplets $(a_i,p_i,t_i)$, referring to health, productivity and lifetime, respectively, for each individual $i$, and we consider a variety of evaluation functions, with different functional forms. Common to most of them is to include a power ($\gamma$) into the lifetime component of each individual before aggregating. But not in the (quality of) health component or the productivity. That is, our functional forms refer to $t^{\gamma}_i$ rather than $h^{\gamma}_i$, where the latter could be interpreted as $(q(a_i)t_i)^{\gamma}$ or $(p_iq(a_i)t_i)^{\gamma}$. In words, our parameter is reflecting inequality aversion with respect to lifetimes, rather than QALYs (or PQALYs). Consequently, we believe we cannot perform direct comparisons with the empirical estimates from Robson et al. (2024). 

Second, the inclusion of productivity in outcome measures raises ethical and practical concerns. Valuing productivity gains could be seen as favoring interventions for working-age or highly productive populations, potentially at the expense of the retired, disabled, or socioeconomically disadvantaged. This tension between efficiency and equity echoes longstanding concerns in health economics: for instance, conventional QALY maximization has been criticized for ignoring whether benefits accrue to the worst-off (e.g., Whitehead and Ali, 2010). 
While our framework introduces greater flexibility by jointly considering equity and productivity, its application still requires careful normative calibration to prevent productivity weighting from distorting distributive objectives. In this regard, recent analyses of inequality aversion in health preferences highlight how sensitivity to health inequities can guide appropriate parametrization of equity concerns within such models (e.g., Robson et al., 2024, 2025a).\footnote{An important aspect in that context is the distinction between achievement or shortfall inequality (e.g., Erreygers, 2009; Laso de la Vega and Aristondo, 2012).}

We emphasize that in our model both health and productivity are assumed to have maximum levels. 
The assumption on health is consistent with the QALY literature. As for productivity, the assumption is an inherent consequence of measuring it in relative terms. If productivity were measured in absolute terms, upper bounds might not be natural. 
% (see, e.g., Syverson, 2011).
%The former assumption is in line with the QALY literature; The later is natural when it reflect a relative *. If productivity is measured in absolute rather than absolute terms, it may be unbounded (e.g., Syverson, 2011).
%by contrast, the latter might not be a standard assumption in the economic literature on productivity . 
Without an upper bound on productivity, the main results of our paper would likely remain valid, albeit their interpretations, as well as some supporting axioms, should be adjusted. %Several axioms and constructions rely on the existence of a maximal productivity level, which no longer makes sense on an unbounded domain; 
For instance, by reformulating axioms and constructions in terms of monotonicity, or scale properties, rather than resorting to comparisons to a top value (upper bound). %With such adjustments of the core axioms, the representation results themselves largely survive, 
%scale-invariance and transfer axioms still lead to additive or power-type productivity terms, now defined on an unbounded productivity range. However, unbounded productivity makes the curvature restrictions implied by equity axioms more consequential, since without concavity extremely high productivities could dominate aggregate evaluations. 
Overall, allowing for productivity to be unbounded would strengthen the normative role of scale invariance and concavity assumptions (as extremely high productivities could dominate aggregate evaluations), while leaving the core functional forms characterized in the paper essentially unchanged.

Another set of limitations relates to the implementation and data requirements of our model. By extending the evaluations to account for productivity, we implicitly take a broader societal perspective on benefits, akin to counting productivity-adjusted life years.
%\footnote{PALYs have been suggested as a way to quantify disease burden including lost productivity (e.g., Ademi et al., 2021).} 
However, operationalizing such measures in practice demands robust data on patients' employment, caregiving roles, and other productive contributions, which health systems do not always systematically track. Moreover, productivity effects can be context-dependent (varying by labor market conditions and economic settings), introducing uncertainty into evaluations (e.g., Elsler et al., 2017). There is also a risk of double-counting benefits or costs: traditionally, productivity gains are accounted for separately; for instance as cost offsets or indirect benefits in a societal cost-benefit analysis (e.g., Ademi et al., 2021). 
This issue should be avoided by clearly specifying whether productivity effects are incorporated as outcomes within the effectiveness measure or as cost components within the analysis.
%When applied in economic evaluations, to ensure that productivity aspects are not double-counted, by clearly designating each as either an outcome or a cost component.
%Our integrated outcome approach needs to ensure that productivity is not over-weighted when health outcomes already correlate with economic outputs. 
%At the same time, it is important to avoid a potential double counting problem. 
Likewise, if productivity effects are already embedded in the measurement of health states, for example if health indices incorporate work outcomes or income consequences, then adding productivity as a separate argument in the social evaluation may count the same underlying effect twice. Our framework therefore presumes a clear conceptual distinction. Health states should capture intrinsic aspects of well-being related to health, while productivity should reflect instrumental contributions to output or income. When this separation is respected, modeling health and productivity as distinct attributes clarifies rather than conflates their respective normative roles. 
%In empirical investigations to elicit health state valuations and productivity, focus should likewise be on limiting double counting for example by asking respondents not to include `work ability' when valuing health dimensions like `mobility' or `usual activities', as the former will be considered separately.

One of our core axioms was \textit{positive lifetime desirability}, which rules out health states that are socially evaluated as worse than zero lifetime.
%implied that all health states are worth living. 
This is at odds with much of the Health-Related Quality of Life (HRQoL) literature (including the EQ-5D we cite in our examples). Now, extending our analysis by dismissing this axiom to account for health states that are worse than death presents conceptual difficulties that motivated our choice. First, the notion of healthy productive years equivalent (which is a central pillar of our analysis) would no longer be well defined. Second, this would bring considerations on population ethics to our analysis, which would require a variable-population version of our model. We can nevertheless stress that some of the functional forms we characterize in this paper could indeed be extended to account for health states that are worse than death (think, for instance, of the equity-sensitive power QALY evaluation function in which quality weights are allowed to be negative). But the normative foundations for them would require a rather different approach from the one we have in this paper (which relies on the concept of healthy productive years equivalent). 

Finally, our model is developed in a static setting, evaluating distributions of health and productivity at a given point or over a fixed interval. This static perspective may oversimplify issues like life-course equity; for instance, how to value equal-sized health gains across different ages (e.g., Tsuchiya, 2000; Bognar, 2008)
%the “fair innings” problem of how to value gains at different ages (e.g., Adler, 2012) 
and long-term dynamic effects on the workforce. These simplifications were necessary to obtain tractable characterizations, but they limit the immediate applicability of the model to longitudinal or intergenerational policy questions.

\subsection{Future research directions} 
%We believe that our findings open up several promising avenues for further research. 
Beyond the conceptual and practical limitations listed above, several avenues remain open for further research.

A natural direction would be to refine the equity-weighting scheme by drawing on empirical social preferences. Future studies could use stated-preference methods or participatory deliberations to pin down society's trade-offs between equity, health and productivity objectives. For example, building on the approach of Lancsar et al. (2011), who derived distributional QALY weights via discrete-choice experiments, one could elicit how much productivity gain the public is willing to forego to favor disadvantaged groups in health allocations. Such research would help translate the theoretical evaluation functions into practical decision rules consistent with societal values. A related empirical question is how decision-makers in practice might respond to an integrated metric. While equity-weighted QALYs and distributional cost-effectiveness analysis have been discussed for some time (e.g., Culyer and Wagstaff, 1993; Williams and Cookson, 2000), agencies have only cautiously begun to incorporate these concepts (e.g., Avance\~na and Prosser, 2021). NICE in the UK is one such example, moving in that direction (e.g., NICE, 2025). Understanding the institutional and political barriers to adopting equity- and productivity-adjusted metrics will be critical for policy impact.

Another important direction would be to extend our framework to capture different dimensions of equity and to test its implications in diverse settings. Equity in health can concern many attributes, not just the level of health but also age, baseline risk, or cause of disadvantage. Recent work has started to disentangle aversion to different sources of health inequality, such as pure health inequality vs. income-related health gaps (e.g., Robson et al., 2024), and to incorporate responsibility-sensitive equity weights that could also be extended to productivity differences (e.g., Robson et al., 2025a). Incorporating this into our model would likely require multi-attribute equity weights or more complex social welfare functions, but it could make the approach more responsive to policy-relevant ethical criteria (such as priority to the worse-off or to those with illness not caused by their choices). Similarly, the relationship between health improvements and productivity may vary by context (consider interventions in an aging society versus a workforce-focused occupational health program). Future research could explore context-specific versions of our evaluation function, potentially drawing on the equity-sensitive compromise measures of Section 4 to tailor the balance between health and productivity to different settings. 
There is also scope to link our model with global health equity analyses. In low- and middle-income settings, productivity gains from health interventions (through improved capacity to work or attend school) are often central to their value (e.g., Premji and Griffin, 2024). Embedding our approach in evaluations for such contexts could provide a more comprehensive assessment of interventions targeting both health and economic development.

A further natural extension of the framework would allow health and productivity to vary over the lifetime. One possible approach would be to model individuals by trajectories $(a(s),p(s))_{s\in[0,t]}$ and to define HPYEs through equivalence between such trajectories and benchmark trajectories at full health and productivity. Under suitable separability or path-independence assumptions, this could lead to HPYE measures that aggregate instantaneous health and productivity over time before applying the equity-sensitive social evaluation developed here. Such a dynamic extension would permit analysis of interventions that affect not only total lifetime but also the timing and evolution of health and productivity, and would raise new questions concerning within-lifetime versus between-individual equity. Exploring the axiomatic foundations of these dynamic HPYE representations is left for future research.

Finally, further theoretical work could strengthen the normative foundations of integrating equity and productivity. Our characterizations were based on axioms that naturally extend standard welfare-economic principles to an outcome space defined by health and productivity.
Nonetheless, additional axioms or alternative formulations might be considered. For instance, one could require consistency with lifetime utility egalitarianism or incorporate elements of prioritarianism (e.g., Robson et al., 2024), ensuring, in our context, that improvements for those who are worse off in \textit{both} health and productivity receive the highest priority. Recent contributions in the literature offer useful ideas to develop further. To name a few, Da Costa et al. (2024) propose a distribution-sensitive health index that could potentially be merged with our productivity-adjusted measure, and Adler et al. (2021) develop a welfare-theoretic interpretation of the fair innings principle over uncertain lifetimes.\footnote{See also Adler et al. (2025).} %Integrating such ideas may produce an even richer framework for evaluations. 
We also note that our model could be extended to incorporate monetary dimensions alongside health, akin to an analysis that integrates quality of life, life years and productivity in a common evaluation function. Bridging cost-benefit analysis and cost-effectiveness analysis in this way (e.g., Bleichrodt and Quiggin, 1999; Herrera-Araujo et al., 2020) is another natural extension suggested by our analysis.

\section{Appendix: Proofs of the results}
\subsection*{CORE axioms}
We start by providing the formal definitions of the CORE axioms, which were introduced and described by their full names in Section 2.2.

\vskip3mm\noindent \textbf{ANON}: $(d_1,...,d_n)\sim (d_{\pi(1)},...,d_{\pi(n)})$ for each $d\in D$, and each bijection $\pi: N\to N$.
%\medskip

\vskip3mm\noindent \textbf{SEP}: For each pair $d,d'\in D$, and each $S\subset N$, $(d_S,d_{N\setminus S}) \succsim (d'_S,d_{N\setminus S}) \Leftrightarrow (d_S,d'_{N\setminus S}) \succsim (d'_S,d'_{N\setminus S})$. 
%\medskip

\vskip3mm\noindent \textbf{CONT}: Let $d,d'\in D$, and $d^{(k)}$ be a sequence such that for each $i \in N$, $d_i^{(k)}=(a_i,p_i^{(k)}, t_i^{(k)}) \rightarrow (a_i,p_i, t_i)=d_i$. If $d^{(k)} \succsim d'$ for each $k$ then $d \succsim d'$, and if $d' \succsim d^{(k)}$ for each $k$, then $d' \succsim d$. 
%\medskip

\vskip3mm\noindent \textbf{ZERO}: For each $d \in D$ and each $i \in N$ such that $t_i = 0$, for all $a^{\prime}_{i}\in A$ and $p_i' \in [0,1]$, $d\sim [d_{N\setminus \{i\}}, (a_i', p_i', 0)]$. 
%\medskip

\vskip3mm\noindent \textbf{FHPS}: For each $d\in D$ and each $i\in N$, $[(a_\ast,p_i, t_i),d_{N\setminus \{i\}}] \succsim d$ and $[(a_i,1, t_i),d_{N\setminus \{i\}}] \succsim d$.
%\medskip

\vskip3mm\noindent \textbf{LMFHP}: For each $d \in D$ and each $i \in N$, such that $(a_i,p_i)=(a_\ast,1)$ and each $t_i > t_i'$, $[(a_\ast,1, t_i), d_{ N\setminus \{i\}}] \succ [(a_\ast,1,t_i'), d_{ N\setminus \{i\}}]$. 
%\medskip

\vskip3mm\noindent \textbf{PLD}: For each $d \in D$ and each $i \in N$, $ d \succsim [(a_i,p_i,0), d_{N\setminus \{i\}}]$. 
%\medskip

\vskip3mm\noindent \textbf{PDTFHP}: For each $d=[(a_{\ast},1,t_{k})_{k\in N}]$, and each pair $i,j\in N$, such that $t_{i}< t_{j}$, and each $0<\tau<\frac{t_{j}-t_{i}}{2}$, %$h^{\prime }=[(a_{\ast}, t^{\prime }_{i})_{i\in N}]$
\begin{equation*}
\left[ \left(a_{\ast},1,t_{i}+\tau\right), \left(a_{\ast},1,t_{j}-\tau\right),d_{N\backslash \{i,j\}}\right]\succ d.
\end{equation*}

\subsection*{Proof of Theorem \ref{thm:genHPYE}}
%\begin{proof}
    By Theorem 10 in Hansen et al. (2024), which draws on Debreu's (1960) classical result on additive representations of separable preferences, generalized HPYE evaluation functions are characterized by the basic structural assumptions, i.e., all the axioms of the CORE pack, except for PDTFHP. Thus, it only remains to show that only those generalized HPYEs with a strictly concave $g$ function satisfy PDTFHP. 
    
    Let $d=[(a_{\ast},1,t_{k})_{k\in N}]$, $i,j\in N$ such that $t_{i}< t_{j}$, and $0<\tau<\frac{t_{j}-t_{i}}{2}$. Then, 
    \begin{equation*}
        \left[ \left(a_{\ast},1,t_{i}+\tau\right), \left(a_{\ast},1,t_{j}-\tau\right),d_{N\backslash \{i,j\}}\right]
        \succ d,
        \end{equation*}
if and only if 
    $$g\left(t_{i}+\tau\right)+g\left(t_{j}-\tau\right)=g\left(f(a_{\ast},1,t_{i}+\tau)\right)+g\left(f(a_{\ast},1,t_{j}-\tau)\right) > g(f(a_{\ast},1,t_{i}))+g(f(a_{\ast},1,t_{j}))=g(t_{i})+g(t_{j}),$$  
    %where $g: \mathbb{R}_{+} \rightarrow \mathbb{R}$ is a strictly increasing, concave and continuous function, and $ f :A\times [0,1]\times \Bbb{R}_+\rightarrow \mathbb{R}_+$ is continuous with respect to its second and third variables and for each $d=[d_1,\dots,d_n]=[(a_1,p_1,t_1),\dots, (a_n,p_n,t_n)] \in D$, 
    %$$d \sim [(a_\ast,1, f(a_i,p_i, t_i))_{i\in N}].$$
    %where, for each $(a_i,p_i,t_i) \in A\times [0,1]\times \Bbb{R}_+$, $0 \le f(a_i,p_i,t_i) \le t_i$,  $f(a_i,p_i,t_i) \le f(a_i,1,t_i)$, and $f(a_i,p_i,t_i) \le f(a_\ast,p_i,t_i)$.
%Equivalently, if and only if 
%$$g\left(\frac{t_{i}+t_{j}}{2}\right)\ge \frac{1}{2}\left(g(t_{i})+g(t_{j})\right),$$
which is equivalent to requiring that $g$ is a strictly concave function, as desired. 
\endproof
 
%\end{proof}   
\subsection*{Proof of Theorem \ref{thm:powerf}}

%\begin{proof}
    Suppose first that $\succsim $ is represented by an evaluation function satisfying $\eqref{gammaf}$. As this is a special case of \eqref{eqgenHPYE}, it follows from Theorem \ref{thm:genHPYE} that CORE holds. 
    As for TSIFHP, let $c>0$, and $d=[(a_*,1,t_1),\dots, (a_*,1,t_n)]\in D,d^{\prime}=[(a_*, 1, t'_1),\dots, (a_*,1,t'_n) ]\in D$. Then,
    $$E^{\gamma f}[d]= E^{\gamma f}[(a_\ast,1,t_1),\dots, (a_\ast,1,t_n)]=\sum_{i=1}^n f(a_\ast,1,t_i)^{\gamma}=\sum_{i=1}^n t_i^{\gamma},$$
    and
    $$E^{\gamma f}[d']= E^{\gamma f}[(a_\ast,1, t_1'),\dots, (a_\ast,1,t'_n) ]=\sum_{i=1}^n t_i'^{ \gamma}.$$
    Thus, 
    $$d\succsim d^{\prime} \iff \sum_{i=1}^n t_i^{\gamma}\ge \sum_{i=1}^n t_i'^{ \gamma}.$$ Similarly, 
    $$
    [(a_*,1,ct_1),\dots, (a_*,1,ct_n)]\succsim [(a_*, 1, ct'_1),\dots, (a_*,1,ct'_n)]\iff c^{\gamma}\sum_{i=1}^n t_i^{\gamma}\ge c^{\gamma}\sum_{i=1}^n t_i'^{\gamma}.$$

    Conversely, assume now that preferences satisfy all the axioms in the statement of Theorem \ref{thm:powerf}. Then, by Theorem \ref{thm:genHPYE}, for each pair $d,d'\in D$, 
    $$d\succsim d' \Leftrightarrow \sum_{i=1}^n g(f(a_i,p_i,t_i))\ge \sum_{i=1}^n g(f(a_i',p_i',t_i')),$$ 
    where $g: \mathbb{R}_{+} \rightarrow \mathbb{R}$ is a strictly increasing, concave and continuous function, and $ f :A\times [0,1]\times \Bbb{R}_+\rightarrow \mathbb{R}_+$ is continuous with respect to its second and third variables such that for each $(a_i,p_i,t_i) \in A\times [0,1]\times \Bbb{R}_+$, $0 \le f(a_i,p_i,t_i) \le t_i$,  $f(a_i,p_i,t_i) \le f(a_i,1,t_i)$, and $f(a_i,p_i,t_i) \le f(a_\ast,p_i,t_i)$, and for each $d=[d_1,\dots,d_n]=[(a_1,p_1,t_1),\dots, (a_n,p_n,t_n)] \in D$, 
    $$d \sim [(a_\ast,1, f(a_i,p_i, t_i))_{i\in N}].$$
    
    By TSIFHP,
    \begin{equation*}
    \sum_{i=1}^{n}g\left( f(a_*,1,t_{i})\right) \geq \sum_{i=1}^{n}g\left(
    f(a_{*},1,t_{i}^{\prime})\right) \iff \sum_{i=1}^{n}g\left(
    cf(a_{*},1,t_{i})\right) \geq \sum_{i=1}^{n}g\left( c f(a_{*},1,t_{i}^{\prime })\right) ,
    \end{equation*}%
    for each $d=[(a_{*},1,t_{1}),\dots ,(a_{*},1,t_{n})]\in D$, $d^{\prime
    }=[(a_{*},1,t_{1}^{\prime}),\dots ,(a_{*},1,t_{n}^{\prime
    })]\in D$ and $c>0$.
    
    %By Bergson and Samuelson (e.g., Burk, 1936; Samuelson, 1965), as in the previous proof, it follows that $E=E^{\gamma f}$, as desired. 
    By Bergson and Samuelson (e.g., Burk, 1936; Samuelson, 1965), there are only
    three possible functional forms for an evaluation function $E$ representing these preferences, when restricted to the domain of distributions with strictly positive lifetimes:
    
    \begin{itemize}
    \item $E[d_{1},\dots,d_{n}]=E[(a_{1},p_1,t_{1}),\dots,(a_{n},p_n,t_{n})]=
    \sum_{i=1}^{n}\alpha_{i}\left(f(a_i,p_i,t_i)\right)^{\gamma},$
    
    \item $E[d_{1},\dots,d_{n}]=E[(a_{1},p_1,t_{1}),\dots,(a_{n},p_n,t_{n})]=
    -\sum_{i=1}^{n}\alpha_{i}\left(f(a_i,p_i,t_i)\right)^{\delta},$
    
    \item $E[d_{1},\dots,d_{n}]=E[(a_{1},p_1,t_{1}),\dots,(a_{n},p_n,t_{n})]=
    \sum_{i=1}^{n}\alpha_{i}\log\left(f(a_i,p_i,t_i)\right),$
    \end{itemize}
    \noindent for some $\gamma >0$, $\delta <0$ and $%
    \alpha _{i}>0$ for each $i\in N$.
    
    It is straightforward to show that the last two functional forms cannot be
    continuously extended to the whole domain $D$, in which zero lifetimes are
    allowed. By ANON, which is part of CORE, $\alpha _{i}=\alpha _{j}$ for each pair $i,j\in N$. By PDTFHP, which is part of CORE, $0<\gamma<1$. Thus, it follows that $E=E^{\gamma f}$, as desired. %Finally, let $\varphi:A\times[0,1]\rightarrow \mathbb{R}$ be such that $\varphi(a,p)=\bar{\varphi}(a,p)^{\gamma }$, for each $(a,p)\in A\times[0,1]$. Altogether, we have that $E=E^{\gamma\varphi}$, as desired.  
    \endproof
  %  \end{proof}

  \subsection*{Proof of Theorem \ref{thm:power}}
  %\begin{proof}
    Suppose first that $\succsim $ is represented by an evaluation function satisfying $\eqref{gamma}$. As this is a special case of \eqref{eqgenHPYE}, with $g(x)=x^{\gamma}$ and $f(a_i,p_i,t_i)=\left(\varphi(a_i,p_i)\right)^{\frac{1}{\gamma}}t_i$, it follows from Theorem \ref{thm:genHPYE} that CORE holds. 
    As for TSI, let $c>0$, and $d=[(a_1,p_1,t_1),\dots, (a_n,p_n,t_n)]\in D,d^{\prime}=[(a_1', p_1', t_1'),\dots, (a'_n,p'_n,t'_n) ]\in D$. Then,
    $$E^{\gamma\varphi}[d]= E^{\gamma\varphi}[(a_1,p_1,t_1),\dots, (a_n,p_n,t_n)]=\sum_{i=1}^n \varphi(a_i,p_i)t_i^{\gamma},$$
    and
    $$E^{\gamma\varphi}[d']= E^{\gamma\varphi}[(a_1', p_1', t_1'),\dots, (a'_n,p'_n,t'_n) ]=\sum_{i=1}^n \varphi(a'_i,p'_i)t_i'^{\gamma}.$$
    Thus, 
    $$d\succsim d^{\prime} \iff \sum_{i=1}^n \varphi(a_i,p_i)t_i^{\gamma}\ge \sum_{i=1}^n \varphi(a'_i,p'_i)t_i'^{\gamma}.$$ Similarly, 
    \begin{equation*}
    \begin{split}
    &[(a_1,p_1,ct_1),\dots, (a_n,p_n,ct_n)]\succsim [(a_1', p_1', ct'_1),\dots, (a'_n,p'_n,ct'_n)] \iff\\
    & c^{\gamma}\sum_{i=1}^n \varphi(a_i,p_i)t_i^{\gamma}\ge c^{\gamma}\sum_{i=1}^n \varphi(a'_i,p'_i)t_i'^{\gamma} \iff\\
    & d\succsim d^{\prime}.
    \end{split}
    \end{equation*}
    
    Conversely, assume now that preferences satisfy all the axioms in the statement of Theorem \ref{thm:power}.\footnote{The ensuing proof extends that of Theorem 4 in Hougaard et al. (2013).} Then, by Theorem \ref{thm:genHPYE}, for each pair $d,d'\in D$, 
    $$d\succsim d' \Leftrightarrow \sum_{i=1}^n g(f(a_i,p_i,t_i))\ge \sum_{i=1}^n g(f(a_i',p_i',t_i')),$$ 
    where $g: \mathbb{R}_{+} \rightarrow \mathbb{R}$ is a strictly increasing, concave and continuous function, and $ f :A\times [0,1]\times \Bbb{R}_+\rightarrow \mathbb{R}_+$ is continuous with respect to its second and third variables, such that for each $(a_i,p_i,t_i) \in A\times [0,1]\times \Bbb{R}_+$, $0 \le f(a_i,p_i,t_i) \le t_i$,  $f(a_i,p_i,t_i) \le f(a_i,1,t_i)$, and $f(a_i,p_i,t_i) \le f(a_\ast,p_i,t_i)$, and for each $d=[d_1,\dots,d_n]=[(a_1,p_1,t_1),\dots, (a_n,p_n,t_n)] \in D$, 
    $$d \sim [(a_\ast,1, f(a_i,p_i, t_i))_{i\in N}].$$
    
    \textbf{Step 1}. We claim that for each pair $(a,p,t),(a^{\prime
    },p^{\prime },t^{\prime })\in A\times[0,1]\times \Bbb{R}_+$, and each $c>0$,
    \begin{equation*}
    f(a,p,t)\geq f(a^{\prime },p^{\prime },t^{\prime })\iff
    f(a,p,ct)\geq f(a^{\prime },p^{\prime },ct^{\prime }).
    \end{equation*}%
    Indeed, let $d=[(a_{1},p_1,t_{1}),\dots ,(a_{n},p_n,t_{n})]\in D$ and $c>0$. Denote $%
    d^{c}=[(a_{1},p_1,ct_{1}),\dots ,(a_{n},p_n,ct_{n})]$ and let $%
    (a,p,t),(a^{\prime },p^{\prime },t^{\prime })\in A\times[0,1]\times \Bbb{R}_+$. By SEP, which is part of CORE,
    \begin{equation*}
    [(a,p,t),d_{N\setminus \{i\}}]\succsim  [(a^{\prime
    },p^{\prime },t^{\prime }),d_{N\setminus \{i\}}]\iff f(a,p,t)\geq
    f(a^{\prime },p^{\prime },t^{\prime }),
    \end{equation*}%
    and
    \begin{equation*}
    [ (a,p,ct),d_{N\setminus \{i\}}^{c}]\succsim [
    (a^{\prime },p^{\prime },ct^{\prime }),d_{N\setminus \{i\}}^{c}]\iff
    f(a,p,ct)\geq f(a^{\prime },p^{\prime },ct^{\prime }).
    \end{equation*}%
    By TSI,
    \begin{equation*}
    [ (a,p,t),d_{N\setminus \{i\}}]\succsim [ (a^{\prime
    },p^{\prime },t^{\prime }),d_{N\setminus \{i\}}]\iff [
    (a,p,ct),d_{N\setminus \{i\}}^{c}]\succsim [ (a^{\prime
    },p^{\prime },ct^{\prime }),d_{N\setminus \{i\}}^{c}].
    \end{equation*}%
    Combining these equivalences concludes. \medskip
    
    \textbf{Step 2}. We now claim the following. Let $\bar{\varphi}:A\times[0,1]\rightarrow
    \mathbb{R}$ be such that $\bar{\varphi}(a,p)=f(a,p,1)$, for each $(a,p)\in A\times[0,1]$.
    Then,
    \begin{equation*}
    f(a,p,t)\geq f(a^{\prime },p^{\prime },t^{\prime })\iff \bar{\varphi}%
    (a,p)t\geq \bar{\varphi}(a^{\prime },p^{\prime })t^{\prime },
    \end{equation*}%
    for each pair $(a,p,t),(a^{\prime },p^{\prime },t^{\prime })\in A\times[0,1]\times \Bbb{R}_{++}$.
    
    %Without loss of generality, we can assume $t\ne 0\ne t'$ (as, otherwise, the statement is trivial). 
    Indeed, let $t,t'\in\Bbb{R}_{++}$. By definition, $f(a,p,1)=\bar{\varphi}(a,p)=f(a_{\ast },1,\bar{\varphi}(a,p))$. By Step 1,
    \begin{equation*}
    f(a,p,t) \geq f(a^{\prime },p^{\prime },t^{\prime })\iff
    f(a,p,ct) \geq f(a^{\prime },p^{\prime },ct^{\prime }).
    \end{equation*}%
    %Thus, $f(a_{i},p_i,t_{i})=\bar{\varphi}(a_{i},p_i)t_{i}$, as desired.
    
    Thus, 
    \begin{equation*}
    \begin{split}
        &f(a,p,t) \geq f(a^{\prime },p^{\prime },t^{\prime })\iff \\
        &f(a,p,1) \geq f(a^{\prime },p^{\prime },\frac{t^{\prime }}{t}) \iff \\
      &f(a_{\ast },1,\bar{\varphi}(a,p)) \geq f(a^{\prime },p^{\prime },\frac{t^{\prime }}{t}) \iff \\
    &f(a_{\ast },1, \frac{t}{t^{\prime }} \bar{\varphi}(a,p)) \geq f(a^{\prime },p^{\prime },1)=\bar{\varphi}(a',p')=f(a_*,1,\bar{\varphi}(a',p')) \iff \\
    &t\bar{\varphi}(a,p) \geq t'\bar{\varphi}(a',p'),
      \end{split}
    \end{equation*}
    as desired.
    
    \textbf{Step 3}. We claim there exists $0<\gamma<1$ such that, for each $%
    d=[(a_{1},p_1,t_{1}),\dots,(a_{n},p_n,t_{n})]\in D$, and each $d^{\prime}=[(a^{%
    \prime}_{1},p^{\prime }_1,t^{\prime}_{1}),\dots,(a^{\prime}_{n},p^{\prime }_n,t^{\prime}_{n})]\in D$,
    \begin{equation*}
    d\succsim d^{\prime} \iff \sum_{i=1}^{n} \left(\bar{\varphi}(a_{i},p_i)t_{i}\right)^{%
    \gamma}\ge \sum_{i=1}^{n} \left(\bar{\varphi}(a^{\prime}_{i},p^{\prime }_i)t^{\prime}_{i}%
    \right)^{\gamma}.
    \end{equation*}
    By Step 2, $f(\cdot ,\cdot,\cdot )$ is a strictly monotonic transformation of the function $%
    \tau :A\times[0,1]\times \Bbb{R}_+\rightarrow \mathbb{R}$ defined by $\tau (a,p,t)=\bar{\varphi}%
    (a,p)t$, for each $(a,p,t)\in A\times[0,1]\times \Bbb{R}_+$. 
    Thus, by Theorem \ref{thm:genHPYE}, there is an increasing function $\bar{g}$ such that the evaluation function defined by 
    \begin{equation*}
    E[d_{1},\dots,d_{n}]=E[(a_{1},p_1,t_{1}),\dots,(a_{n},p_n,t_{n})]=\sum_{i=1}^{n}
    \bar{g}\left(\bar{\varphi}(a_{i},p_i)t_{i}\right),
    \end{equation*}
    represents $\succsim. $
    
    %Then, by SEP, $E$ represents $\succsim $. 
    
    By TSI,
    \begin{equation*}
    \sum_{i=1}^{n}\bar{g}\left( \bar{\varphi}(a_{i},p_i)t_{i}\right) \geq \sum_{i=1}^{n}\bar{g}\left(
    \bar{\varphi}(a_{i}^{\prime },p_{i}^{\prime })t_{i}^{\prime }\right) \iff \sum_{i=1}^{n}\bar{g}\left(
    \bar{\varphi}(a_{i},p_i)ct_{i}\right) \geq \sum_{i=1}^{n}\bar{g}\left( \bar{\varphi}%
    (a_{i}^{\prime },p_{i}^{\prime })ct_{i}^{\prime }\right) ,
    \end{equation*}%
    for each $d=[(a_{1},p_1,t_{1}),\dots ,(a_{n},p_n,t_{n})]\in D$, each $d^{\prime
    }=[(a_{1}^{\prime },p_{1}^{\prime },t_{1}^{\prime }),\dots ,(a_{n}^{\prime },p_{n}^{\prime },t_{n}^{\prime
    })]\in D$ and each $c>0$.
    
    By Bergson and Samuelson (e.g., Burk, 1936; Samuelson, 1965), as in the previous proof, there are only
    three possible functional forms for $E$, when restricted to the domain of distributions with strictly positive lifetimes:
    
    \begin{itemize}
    \item $E[d_{1},\dots,d_{n}]=E[(a_{1},p_1,t_{1}),\dots,(a_{n},p_n,t_{n})]=
    \sum_{i=1}^{n}\alpha_{i}\left(\bar{\varphi}(a_{i},p_i)t_{i}\right)^{\gamma},$
    
    \item $E[d_{1},\dots,d_{n}]=E[(a_{1},p_1,t_{1}),\dots,(a_{n},p_n,t_{n})]=
    -\sum_{i=1}^{n}\alpha_{i}\left(\bar{\varphi}(a_{i},p_i)t_{i}\right)^{\delta},$
    
    \item $E[d_{1},\dots,d_{n}]=E[(a_{1},p_1,t_{1}),\dots,(a_{n},p_n,t_{n})]=
    \sum_{i=1}^{n}\alpha_{i}\log\left(\bar{\varphi}(a_{i},p_i)t_{i}\right),$
    \end{itemize}
    \noindent for some $\gamma >0$, $\delta <0$ and $%
    \alpha _{i}>0$ for each $i\in N$.
    
    It is straightforward to show that the last two functional forms cannot be
    continuously extended to the whole domain $D$, in which zero lifetimes are
    allowed. By ANON, which is part of CORE, $\alpha _{i}=\alpha _{j}$ for each pair $i,j\in N$. By PDTFHP, which is part of CORE, $0<\gamma<1$. Finally,
    let $\varphi:A\times[0,1]\rightarrow \mathbb{R}$ be such that $\varphi(a,p)=\bar{\varphi}%
    (a,p)^{\gamma }$, for each $(a,p)\in A\times[0,1]$. Altogether, we have that $E=E^{\gamma\varphi}
    $, as desired. \endproof
    %\end{proof}
    \medskip

\subsection*{Proof of Proposition \ref{power-QALY}}
    Suppose first that $\succsim $ is represented by an evaluation function satisfying $(\ref{power-eqQALY})$. As this is a special case of $\eqref{gamma}$, it follows by Theorem \ref{thm:power} that it satisfies CORE and TSI. It is straightforward to see that it also satisfies PI. 
    
    Conversely, assume now that preferences satisfy all the axioms in the statement of Proposition \ref{power-QALY}. Then, by Theorem \ref{thm:power}, for each pair $d,d'\in D$, 
    $$d\succsim d' \Leftrightarrow \sum_{i=1}^n \varphi(a_i,p_i)t_i^{\gamma}\ge \sum_{i=1}^n \varphi(a'_i,p'_i)t_i'^{\gamma},$$ 
    where $\varphi:A\times [0,1] \rightarrow [0,1]$ is a continuous function in its second variable such that $0\le \varphi(a,p)\le \varphi(a_{\ast},1)=1$, $\varphi(a,p) \le \varphi(a_{\ast},p),$ and $\varphi(a,p) \le \varphi(a,1)$ for each $(a,p)\in A\times[0,1]$, and $0<\gamma <1$. By PI, $\varphi(a,p)=\varphi(a,p')$ for each pair $p,p'\in [0,1]$. Thus, let $q:A \rightarrow [0,1]$ be such that $q(a)=\varphi(a,1)$, for each $a\in A$. It follows that $0\le q(a)\le q(a_{\ast})=1$ and that for each pair $d,d'\in D$, 
    $$d\succsim d' \Leftrightarrow \sum_{i=1}^n q(a_i)t_i^{\gamma}\ge \sum_{i=1}^n q(a'_i)t_i'^{\gamma},$$
    which concludes the proof. \endproof
    \subsection*{Proof of Proposition \ref{power-PALY}}

    Suppose first that $\succsim $ is represented by an evaluation function satisfying $(\ref{power-eqPALY})$. As this is a special case of $\eqref{gamma}$, it follows by Theorem \ref{thm:power} that it satisfies CORE and TSI. It is straightforward to see that it also satisfies the remaining axioms. 
    
    Conversely, assume now that preferences satisfy all the axioms in the statement of Proposition \ref{power-PALY}. Then, by Theorem \ref{thm:power}, for each pair $d,d'\in D$, 
    $$d\succsim d' \Leftrightarrow \sum_{i=1}^n \varphi(a_i,p_i)t_i^{\gamma}\ge \sum_{i=1}^n \varphi(a'_i,p'_i)t_i'^{\gamma},$$ 
    where $\varphi:A\times [0,1] \rightarrow [0,1]$ is a continuous function in its second variable such that $0\le \varphi(a,p)\le \varphi(a_{\ast},1)=1$, $\varphi(a,p) \le \varphi(a_{\ast},p),$ and $\varphi(a,p) \le \varphi(a,1)$ for each $(a,p)\in A\times[0,1]$, and $0<\gamma <1$. By HI, $\varphi(a,p)=\varphi(a',p)$ for each pair $a,a'\in A$. Thus, let $v:[0,1] \rightarrow [0,1]$ be such that $v(p)=\varphi(a_{\ast},p)$, for each $p\in [0,1]$. It follows that $0\le v(p)\le v(1)=1$ and that for each pair $d,d'\in D$, 
    $$d\succsim d' \Leftrightarrow \sum_{i=1}^n v(p_i)t_i^{\gamma}\ge \sum_{i=1}^n v(p'_i)t_i'^{\gamma}.$$ 
    Now, by PICHT, it follows that $v(p_i+c)-v(p_i)=v(p_j+c)-v(p_j)$, for each pair $p_i,p_j\in [0,1]$ and each $c>0$ such that $p_i+c,p_j+c\in [0,1]$. In particular, 
$$
v\left(\frac{x+y}{2}\right)=\frac{v(x)+v(y)}{2},
$$
for each $x,y\in [0,1]$. %\footnote{Take $t_{i}=x$, $c=\frac{y-x}{2}$, and $t_{j}=y-c=\frac{y+x}{2}$.} 
As $v$ is continuous and bounded, it follows from Theorem 1 in Aczél (2006, p. 43) that there exist $\alpha,\beta\in \mathbb{R}$  such that $v(x)=\alpha x+\beta$, for each $x\in [0,1]$. By TIUP, $v(0)=0$, and thus $\beta=0$. Thus, $1=v(1)=\alpha$. Consequently, $\succsim $ is indeed represented by an evaluation function satisfying $(\ref{power-eqPALY})$, as desired. \endproof

%\subsection*{Proofs of the results within Section \ref{Power PQALYs}}
%\begin{proof}
%We focus on the non-trivial implications of each statement. 
\subsection*{Proof of Theorem \ref{thm:power-M-PQALY}}
Suppose first that $\succsim $ is represented by an evaluation function satisfying $(\ref{power-eqM-PQALY})$. As this is a special case of $\eqref{gamma}$, it follows by Theorem \ref{thm:power} that it satisfies CORE and TSI. As for the remaining axioms, it is straightforward to see that (\ref{power-eqM-PQALY}) also satisfies TIUP. 
As for PICHT, let $d\in D$, and $i,j\in N$ with $a_i=a_j=a$ and $t_i=t_j=t$. Then, for each $c>0$ such that $p_i+c,p_j+c \leq 1$,
$$E^{\gamma pq}[(a,p_i+c,t),(a,p_j,t), d_{ N\setminus \{i,j\}}] = q(a)(p_i+c)t^{\gamma}+q(a)p_j t^{\gamma}+\sum_{k\in N\setminus \{i,j\}} q(a_k)p_k t^{\gamma}_k,$$
and
$$E^{\gamma pq}[(a,p_i,t),(a,p_j+c,t), d_{ N\setminus \{i,j\}}]=q(a)p_i t^{\gamma}+q(a)(p_j+c)t^{\gamma}+\sum_{k\in N\setminus \{i,j\}} q(a_k)p_k t^{\gamma}_k.$$
Thus, %As $q(a_\ast)=1$, it follows that 
$$[(a,p_i+c,t),(a,p_j,t), d_{ N\setminus \{i,j\}}]\sim [(a,p_i,t),(a,p_j+c,t), d_{ N\setminus \{i,j\}}].$$
    
Conversely, assume now that preferences satisfy all the axioms in the statement of Theorem \ref{thm:power-M-PQALY}. Then, by Theorem \ref{thm:power}, for each pair $d,d'\in D$, 
    $$d\succsim d' \Leftrightarrow \sum_{i=1}^n \varphi(a_i,p_i)t_i^{\gamma}\ge \sum_{i=1}^n \varphi(a'_i,p'_i)t_i'^{\gamma},$$ 
    where $\varphi:A\times [0,1] \rightarrow [0,1]$ is a continuous function in its second variable such that $0\le \varphi(a,p)\le \varphi(a_{\ast},1)=1$, $\varphi(a,p) \le \varphi(a_{\ast},p),$ and $\varphi(a,p) \le \varphi(a,1)$ for each $(a,p)\in A\times[0,1]$, and $0<\gamma <1$.
    For each $a\in A$, let $\varphi^{a}: [0,1] \rightarrow \mathbb{R}_+$ be such that $\varphi^{a}(p)=\varphi(a,p)$, for each $p\in [0,1]$. Then, $\varphi^{a}$ is a continuous function and, by PICHT, such that $\varphi^{a}(p_i+c)+\varphi^{a}(p_j)=\varphi^{a}(p_i)+\varphi^{a}(p_j+c)$, for each pair $p_i,p_j\in [0,1]$ and each $c>0$ such that $p_i+c,p_j+c\in [0,1]$. In particular, 
    $$
    \varphi^{a}\left(\frac{x+y}{2}\right)=\frac{\varphi^{a}(x)+\varphi^{a}(y)}{2},
    $$
    for each $x,y\in [0,1]$. %\footnote{Take $t_{i}=x$, $c=\frac{y-x}{2}$, and $t_{j}=y-c=\frac{y+x}{2}$.} 
    %As $\varphi^{a}$ is continuous and strictly increasing, it follows from 
    Thus, by Theorem 1 in Aczél (2006, p. 43), %$\varphi^{a}$ is linear.
    there exist $\alpha,\beta\in \Bbb{R}$ such that $\varphi^{a}(x)=\alpha x+\beta$, for each $x\in[0,1]$. %Consequently, there exist $q,r :A\rightarrow\mathbb{R}$ such that $\varphi(a,p)=q(a)p+r(a)$, for each $p\in [0,1]$, and each $a\in A$.
    Consequently, there exist $\alpha,\beta :A\rightarrow\mathbb{R}$ such that $\varphi(a,p)=\alpha(a)p+\beta(a)$, for each $p\in [0,1]$, and each $a\in A$. As $0\le \varphi(a,p)\le \varphi(a_{\ast},p)\le \varphi(a_{\ast},1)=1$, and $0\le \varphi(a,p)\le \varphi(a,1)\le \varphi(a_{\ast},1)=1$, for each $(a,p)\in A\times[0,1]$, it follows that $0\le \alpha(a)p+\beta(a)\le \alpha(a_{\ast})p+\beta(a_{\ast})\le\alpha(a_{\ast})+\beta(a_{\ast})=1$, and $0\le \alpha(a)p+\beta(a)\le \alpha(a)+\beta(a)\le \alpha(a_{\ast})+\beta(a_{\ast})=1$ for each $(a,p)\in A\times[0,1]$. In particular, $\alpha(a)p\le \alpha(a)$, and $0\le \alpha(a)p+\beta(a)$, for each $(a,p)\in A\times[0,1]$, which implies that 
    $1\ge\alpha(a)\ge 0$ and $1\ge\beta(a)\ge 0$, for each $a\in A$. %Also, when $p=0$ in the la$p\in [0,1]$

    Now, by TIUP, it follows that $\varphi(a,0)=\beta(a)=0$ for each $a\in A$. Consequently, if we simply let $q :A\rightarrow\mathbb{R}$ be such that $q(a)=\alpha(a)$, for each $a\in A$, it follows that $\succsim $ is indeed represented by an evaluation function satisfying $(\ref{power-eqM-PQALY})$, as desired. % for the statement of Theorem \ref{thm:power-M-PQALY}.\endproof%\footnote{Note that the monotonicity properties of the $q$ function are obtained from FHPS and, possibly, normalization.}
\endproof
\subsection*{Proof of Theorem \ref{thm:power-QALY-PALY}}
Suppose first that $\succsim $ is represented by an evaluation function satisfying $(\ref{power-QALY-PALY})$. As this is a special case of $\eqref{gamma}$, it follows by Theorem \ref{thm:power} that it satisfies CORE and TSI. As for PICT, %Finally, to show that (\ref{power-QALY-PALY}) satisfies PICT, 
let $d\in D$, and $i,j\in N$ with $t_i=t_j=t$. Then, for each $c>0$ such that $p_i+c,p_j+c \leq 1$,
\begin{equation*}
    \begin{split}
        & E^{\gamma \sigma}[(a_i,p_i+c,t),(a_j,p_j,t), d_{ N\setminus \{i,j\}}] = \\
        & \sigma (q(a_i)t^{\gamma} +q(a_j)t^{\gamma})+(1-\sigma)((p_i+c)t^{\gamma}+p_j t^{\gamma})+\sum_{k\in N\setminus \{i,j\}} \left( \sigma q(a_k)t^{\gamma}_k+(1-\sigma) p_kt^{\gamma}_k \right),
    \end{split}
    \end{equation*} 
and
\begin{equation*}
    \begin{split}
        & E^{\gamma \sigma}[(a_i,p_i,t),(a_j,p_j+c,t), d_{ N\setminus \{i,j\}}]= \\
        & \sigma (q(a_i)t^{\gamma} +q(a_j)t^{\gamma})+(1-\sigma)(p_i t^{\gamma}+(p_j+c) t^{\gamma})+\sum_{k\in N\setminus \{i,j\}} \left( \sigma q(a_k)t^{\gamma}_k+(1-\sigma) p_kt^{\gamma}_k \right).
\end{split}
\end{equation*} 
Thus, %As $q(a_\ast)=1$, it follows that 
$$[(a_i,p_i+c,t),(a_j,p_j,t), d_{ N\setminus \{i,j\}}]\sim [(a_i,p_i,t),(a_j,p_j+c,t), d_{ N\setminus \{i,j\}}].$$
%The proofs for the remaining results in the section are omitted for being trivial. \endproof
%\medskip  
    
Conversely, assume now that preferences satisfy all the axioms in the statement of Theorem \ref{thm:power-QALY-PALY}. %, i.e., CORE, TSI and PICT. %As PICT implies PICHT, we can replicate the proof of Theorem \ref{thm:power-A-PQALY} to obtain that 
Then, by replicating the argument in the previous proof (up until the last paragraph), we obtain that $\succsim $ is represented by an evaluation function %satisfying $(\ref{power-eqA-PQALY})$, with $\delta\in [0,1]$, as desired for the statement of Theorem \ref{thm:power-A-PQALY}. 
$$\sum_{i=1}^n \left(\alpha(a_i)p_i+\beta(a_i)\right) t_i^{\gamma},$$ 
where %$\varphi:A\times [0,1] \rightarrow [0,1]$ there exist 
$\alpha,\beta :A\rightarrow\mathbb{R}$ are such that %$\varphi(a,p)=\alpha(a)p+\beta(a)$, for each $p\in [0,1]$, and each $a\in A$. As $0\le \varphi(a,p)\le \varphi(a_{\ast},p)\le \varphi(a_{\ast},1)=1$, and $0\le \varphi(a,p)\le \varphi(a,1)\le \varphi(a_{\ast},1)=1$, for each $(a,p)\in A\times[0,1]$, it follows that 
$0\le \alpha(a)p+\beta(a)\le \alpha(a_{\ast})p+\beta(a_{\ast})\le\alpha(a_{\ast})+\beta(a_{\ast})=1$, and $0\le \alpha(a)p+\beta(a)\le \alpha(a)+\beta(a)\le \alpha(a_{\ast})+\beta(a_{\ast})=1$ for each $(a,p)\in A\times[0,1]$. In particular, $\alpha(a)p\le \alpha(a)$, and $0\le \alpha(a)p+\beta(a)$, for each $(a,p)\in A\times[0,1]$, which implies that $1\ge\alpha(a)\ge 0$ and $1\ge\beta(a)\ge 0$, for each $a\in A$. %Also, when $p=0$ in the la$p\in [0,1]$

Now, by PICT, it follows that $\alpha(a_i)(p_i+c)+\alpha(a_{\ast})p_j=\alpha(a_{i})p_i+\alpha(a_{\ast})(p_j+c)$ for each $a_i\in A$, each pair $p_i,p_j\in[0,1]$ and each $c>0$ (such that $\max\{p_i,p_j\}+c\le 1$). Hence, $\alpha(a_{i})=\alpha(a_{\ast})=1-\beta(a_{\ast})$, for each $a_i\in A$. 

Let $\sigma=\beta(a_{\ast})\in [0,1]$. 
We now distinguish several cases.
    
    Case 1. $\sigma=1$. %$\alpha(a_{\ast})=0$. In this case, $\beta(a_{\ast})=1$. Furthermore, 
    %$0\le \alpha(a)p+\beta(a)\le \alpha(a_{\ast})p+\beta(a_{\ast})=1$, and $0\le \alpha(a)p+\beta(a)\le \alpha(a)+\beta(a)\le 1$ for each $(a,p)\in A\times[0,1]$. In particular, $\alpha(a)p\le \alpha(a)$ for each $p\in [0,1]$, which implies that 

    In this case, $\alpha(a)=0$ for each $a\in A$. Therefore, %$\varphi(a,p)=\beta(a)$, for each $(a,p)\in A\times[0,1]$, and 
    $0\le \beta(a)\le \beta(a_{\ast})=1$. If we let $q:A\rightarrow [0,1]$ be such that $q(a)=\beta (a)$, for each $a\in A$, it follows that $\succsim $ is represented by an evaluation function satisfying $(\ref{power-QALY-PALY})$, with $\sigma=1$.
    
    Case 2. $\sigma=0$. 
    
    In this case, $\alpha(a)=1$ for each $a\in A$ and, as $0\le \alpha(a)p+\beta(a)\le p\le 1$ for each $(a,p)\in A\times[0,1]$, it follows that 
    %Furthermore, $0\le \alpha(a)p+\beta(a)\le p\le1$, and $0\le \alpha(a)p+\beta(a)\le \alpha(a)+\beta(a)\le1$ for each $(a,p)\in A\times[0,1]$. Thus, if $p=0$, we obtain 
    $0\le \beta(a)\le 0$, for each $a\in A$. Thus, %$\varphi(a,p)=\alpha(a)p$, for each $(a,p)\in A\times[0,1]$, and  $0\le \alpha(a)\le \alpha(a_{\ast})=1$. if we let $r:A\rightarrow [0,1]$ be such that $r(a)=\alpha (a)$, for each $a\in A$, 
  $\succsim $ is represented by an evaluation function satisfying $(\ref{power-QALY-PALY})$, with $\sigma=0$.
    
    Case 3. $\sigma\in(0,1)$. 
    
    Let $q:A\rightarrow \mathbb{R}$ be such that $q(a)=\frac{\beta(a)}{\beta(a_{\ast})}$, for each $a\in A$. Then, $1=q(a_{\ast})\ge q(a)\ge 0$. Furthermore, we have $\alpha(a)p+\beta(a)=\sigma q(a)+(1-\sigma) p$, for each $p\in [0,1]$, and each $a\in A$. Thus, for each pair $d,d'\in D$, 
    $$d\succsim d' \Leftrightarrow \sigma\sum_{i=1}^n q(a_i)t^{\gamma}_i+(1-\sigma)\sum_{i=1}^n p_i t^{\gamma}_i\ge \sigma\sum_{i=1}^n q(a'_i){t'}^{\gamma}_i+(1-\sigma)\sum_{i=1}^n p'_i{t'}^{\gamma}_i,
    $$
    where $q:A\rightarrow [0,1]$ is such that $1=q(a_{\ast})\ge q(a)\ge 0$, and $\sigma\in (0,1)$. %Consequently, $\succsim $ is indeed represented by an evaluation function satisfying $(\ref{power-eqA-PQALY})$, with $\delta\in (0,1)$. , as desired for the statement of Theorem \ref{thm:power-A-PQALY}.

Thus, $\succsim $ is indeed represented by an evaluation function satisfying $(\ref{power-QALY-PALY})$, as desired. %for the statement of Theorem \ref{thm:power-QALY-PALY}. 
\endproof

\subsection*{Proof of Theorem \ref{thm:power-A-PQALY}}
Suppose first that $\succsim $ is represented by an evaluation function satisfying $(\ref{power-eqA-PQALY})$. As this is a special case of $\eqref{gamma}$, it follows by Theorem \ref{thm:power} that it satisfies CORE and TSI. As for PICHT, %the remaining axioms, %Similarly, for (\ref{power-eqA-PQALY}),
let $d\in D$, and $i,j\in N$ with $a_i=a_j=a$ and $t_i=t_j=t$. Then, for each $c>0$ such that $p_i+c,p_j+c \leq 1$,
\begin{equation*}
\begin{split}
& E^{\gamma rs}[(a,p_i+c,t),(a,p_j,t), d_{ N\setminus \{i,j\}}] = \\ 
& r(a)t^{\gamma} +r(a)t^{\gamma}+s(a)(p_i+c)t^{\gamma}+s(a)p_j t^{\gamma}+ \sum_{k\in N\setminus \{i,j\}} \left( r(a_k)t^{\gamma}_k+ s(a_k)p_kt^{\gamma}_k \right), 
\end{split}
\end{equation*}
and 
\begin{equation*}
\begin{split}
    & E^{\gamma rs}[(a,p_i,t),(a,p_j+c,t), d_{ N\setminus \{i,j\}}]= \\
    & r(a)t^{\gamma} +r(a)t^{\gamma}+s(a)p_i t^{\gamma}+s(a)(p_j+c) t^{\gamma}+\sum_{k\in N\setminus \{i,j\}} \left( r(a_k)t^{\gamma}_k+s(a_k)p_k t^{\gamma}_k \right).
\end{split}
\end{equation*} 
Thus, %As $q(a_\ast)=1$, it follows that 
$$[(a,p_i+c,t),(a,p_j,t), d_{ N\setminus \{i,j\}}]\sim [(a,p_i,t),(a,p_j+c,t), d_{ N\setminus \{i,j\}}].$$ 
    
Conversely, assume now that preferences satisfy all the axioms in the statement of Theorem \ref{thm:power-A-PQALY}. Then, by replicating the argument in the proof of Theorem \ref{thm:power-M-PQALY} (up until the last paragraph), we obtain that $\succsim $ is represented by an evaluation function %satisfying $(\ref{power-eqA-PQALY})$, with $\delta\in [0,1]$, as desired for the statement of Theorem \ref{thm:power-A-PQALY}. 
$$\sum_{i=1}^n \left(\alpha(a_i)p_i+\beta(a_i)\right) t_i^{\gamma},$$ 
where %$\varphi:A\times [0,1] \rightarrow [0,1]$ there exist 
$\alpha,\beta :A\rightarrow\mathbb{R}$ are such that for each $(a,p)\in A\times[0,1]$, %$\varphi(a,p)=\alpha(a)p+\beta(a)$, for each $p\in [0,1]$, and each $a\in A$. As $0\le \varphi(a,p)\le \varphi(a_{\ast},p)\le \varphi(a_{\ast},1)=1$, and $0\le \varphi(a,p)\le \varphi(a,1)\le \varphi(a_{\ast},1)=1$, for each $(a,p)\in A\times[0,1]$, it follows that 
$0\le \alpha(a)p+\beta(a)\le \alpha(a_{\ast})p+\beta(a_{\ast})\le\alpha(a_{\ast})+\beta(a_{\ast})=1$, and $0\le \alpha(a)p+\beta(a)\le \alpha(a)+\beta(a)\le \alpha(a_{\ast})+\beta(a_{\ast})=1$. In particular, for each $(a,p)\in A\times[0,1]$, $0\le \beta(a)\le \beta(a_{\ast})\le 1$, $\alpha(a)p\le \alpha(a)$, and $0\le \alpha(a)p+\beta(a)$, which implies that, for each $a\in A$, $1\ge\alpha(a)\ge 0$ and $1\ge\beta(a)\ge 0$. %Also, when $p=0$ in the la$p\in [0,1]$

Consequently, if we simply let $r :A\rightarrow\mathbb{R}$ be such that $r(a)=\beta(a)$, for each $a\in A$, and $s :A\rightarrow\mathbb{R}$ be such that $s(a)=\alpha(a)$, for each $a\in A$, it follows that $\succsim $ is indeed represented by an evaluation function satisfying $(\ref{power-eqA-PQALY})$, as desired.
\endproof

\subsection*{Proof of Theorem \ref{thm:bi-power-PQALY}}

 Suppose first that $\succsim$ is represented by an evaluation function satisfying $\eqref{gammam-core}$. 
 As this is a special case of \eqref{gamma}, it follows from Theorem \ref{thm:power} that CORE and TSI hold. PDTFHCT follows from the strict concavity of $p^\varepsilon$ for $0 < \varepsilon < 1$.
  
   As for PSI, let $d=[(a_1,p_1,t_1),\dots, (a_n,p_n,t_n)]\in D$, $d^{\prime}=[(a_1', p_1', t'_1),\dots, (a'_n,p'_n,t'_n) ]\in D$ and $0<c<1$. Then,
    $$E^{\gamma \varepsilon}[d]= E^{\gamma \varepsilon}[(a_1,p_1,t_1),\dots, (a_n,p_n,t_n)]=\sum_{i=1}^n q(a_i)p_i^{\varepsilon}t_i^{\gamma},$$
    and
    $$E^{\gamma \varepsilon}[d']= E^{\gamma \varepsilon}[(a_1', p_1', t_1'),\dots, (a'_n,p'_n,t'_n) ]=\sum_{i=1}^n q(a'_i)p_i'^{\varepsilon}t_i'^{\gamma}.$$
    Thus, 
    $$d\succsim d^{\prime} \iff \sum_{i=1}^n q(a_i)p_i^{\varepsilon}t_i^{\gamma}\ge \sum_{i=1}^n q(a'_i)p_i'^{\varepsilon}t_i'^{\gamma}.$$ Similarly, 
    \begin{equation*}
    \begin{split}
    &[(a_1,cp_1,t_1),\dots, (a_n,cp_n,t_n)]\succsim [(a_1', cp_1', t'_1),\dots, (a'_n,cp'_n,t'_n)] \iff\\
    & c^{\varepsilon}\sum_{i=1}^n q(a_i)p_i^{\varepsilon}t_i^{\gamma}\ge c^{\varepsilon}\sum_{i=1}^n q(a'_i)p_i'^{\varepsilon}t_i'^{\gamma} \iff\\
    & d\succsim d^{\prime}.
    \end{split}
    \end{equation*}

  Conversely, assume now that preferences satisfy all the axioms in the statement of Theorem \ref{thm:bi-power-PQALY}. By CORE and TSI, it follows from Theorem \ref{thm:power} that there exists a separable evaluation function of the form (\ref{gamma}). %(3). 

We stress that the Bergson-Samuelson result used in the proofs of the preceding theorems continues to hold in this case (when the domain of the social welfare indicator, over which scale independence is imposed, is restricted to the unit interval $ [0,1]$).\footnote{This follows from the structure of the proof of Theorem 2.6b in Moulin (1988), originally attributed to Roberts (1980), which parallels the Bergson-Samuelson result. Specifically, the derivation involves a Pexider-type functional equation, whose general solutions are either power functions or logarithmic functions. According to Aczél (1987, Corollary 5), such functional forms remain valid solutions even when the functional equation is defined on $[0,1]$ rather than the entire non-negative real line.} %Therefore, the Bergson-Samuelson characterization retains its validity under this restricted domain.}

%Moreover, we note that the Bergson-Samuelson result applies also to separable evaluation functions with given weights. This is shown in Samuelson (1965), p. 787-788, where the weights are given by the constants ``$a_i$'' in the context of his model.

Furthermore, the Bergson-Samuelson result extends to the case of separable evaluation functions with fixed weights.\footnote{This is, for instance, considered by Samuelson (1965, pp. 787-788), where the weights correspond to the constants ``$a_i$'' specified within the framework of his model.}
%\footnote{An alternative way to see this, under slightly stronger assumptions, is found in Bergson’s original formulation (Burk, 1936), where the result is derived via differential equations involving the ratio of the first derivative to the second derivative. The weights do not influence these equations, as they appear as constants in both the numerator and denominator and thus cancel out. The resulting expressions can then be incorporated into the evaluation function using weights. While Bergson did not identify the logarithmic form—this was later observed by Samuelson (1965)—this is immaterial in our context, as our domain includes 0. However, Bergson assumes twice-differentiability, an assumption that is not made in our setting.} 

Thus, by PSI and the Bergson-Samuelson result applied to (\ref{gamma}), considering $t_i^\gamma$ as weights, we get that the evaluation function can be written as
 \begin{equation*}
\sum_{i=1}^n t_i^\gamma\bar{q}{}(a_i) p_i^\varepsilon,
 \end{equation*}
with $\varepsilon>0$. Now, by PDTFHCT, $0<\varepsilon<1$. Finally, by PLD, which is part of CORE, $\bar{q}(a_i) \geq 0$, and by LMFHP, which is also part of CORE, $\bar{q}(a_*) >0$. 
Define $q(a_i)= \frac{\bar{q}(a_i)}{ \bar{q}(a_*)}$. We then get that the evaluation function can be written as 
\begin{equation*}
\
\sum_{i=1}^n q(a_i)p_i^{\varepsilon}t_i^{\gamma},
 \end{equation*}
with $0<\varepsilon,\gamma<1$ as desired. \endproof

\bigskip

%We conclude mentioning that the three evaluation functions in the statements of Theorems \ref{thm:power-M-PQALY}, \ref{thm:power-A-PQALY} and \ref{thm:power-QALY-PALY} are special cases of \eqref{gamma} and, therefore, it follows from Theorem \ref{thm:power} that they all satisfy CORE and TSI. 

%\subsection*{Proof of Theorem \ref{thm:A-PQALY}}
%\begin{proof}
%Suppose first that $\succsim $ is represented by a PHEF satisfying $\eqref{eqA-PQALY}$. As this is a special case of \eqref{power-tw-HPYE}, it follows from Theorem \ref{thm:tw-HPYE} that CORE and TICHP hold. As for PICHT, 

\end{document}